\documentclass[11pt,a4paper]{article} 
\pdfoutput=1
 \usepackage{jheppub}
\usepackage{enumerate}
\usepackage{epsfig} 
\usepackage{amssymb} 
\usepackage{wasysym} 
\usepackage{mathrsfs} 
\usepackage{graphicx} 
\usepackage{amsfonts} 
\usepackage{amsbsy} 
\usepackage{amscd} 
\usepackage{pstricks} 
\usepackage{multirow} 
\usepackage{bm, 
  braket,
  array,
  slashed,
  mathrsfs,
  subfig,
  epsfig
}
\usepackage{color}
\definecolor{myred}{rgb}{0.6,0,0} 
\definecolor{myblue}{rgb}{0,0.2,0.4}
\definecolor{mygreen}{rgb}{0,0.9,0.1}
\definecolor{hc}{rgb}{.9,0.1,0.7}
\definecolor{hcout}{rgb}{.9,0.7,0.9}
\definecolor{Orange}{rgb}{1.,0.65,0.}

\numberwithin{equation}{section}
\numberwithin{figure}{section}
\numberwithin{table}{section}

\newcommand{\be}{\begin{equation}}
\newcommand{\ee}{\end{equation}}
\newcommand{\bea}{\begin{eqnarray}}
\newcommand{\eea}{\end{eqnarray}}

\newcommand{\ba}{\begin{align}}
\newcommand{\ea}{\end{align}}

\def\beq{\begin{eqnarray}}
\def\eeq{\end{eqnarray}}
\def\bea{\begin{eqnarray}}
\def\eea{\end{eqnarray}}

\newcommand{\lsim}{\lesssim}

\newcommand\abs[1]{\left|#1\right|}
\definecolor{blu}{cmyk}{1,0.7,0,0.6}
\definecolor{navyblue}{rgb}{0.0, 0.0, 0.5}
\definecolor{red(ncs)}{rgb}{0.77, 0.01, 0.2}
\definecolor{olive}{rgb}{0.5, 0.5, 0.0}
\definecolor{sapgreen}{rgb}{0.16, 0.95, 0.35}

\title{{\Large The Scalar Triplet Contribution to Lepton Flavour Violation and Neutrinoless Double Beta Decay in Left-Right Symmetric Model}}

 \author[a]{Gulab Bambhaniya,}
\affiliation[a]{Physical Research Laboratory, Navrangpura, Ahmedabad 380009, India}

\author[b,c]{P. S. Bhupal Dev,}
\affiliation[b]{Physik-Department T30d, Technische Univertit\"{a}t M\"{u}nchen, 
James-Franck-Stra\ss e 1, D-85748 Garching, Germany}
\affiliation[c]{Max-Planck-Institut f\"{u}r Kernphysik, Saupfercheckweg 1, D-69117 Heidelberg, Germany}

\author[a]{Srubabati Goswami,}

\author[d]{and Manimala Mitra}
\affiliation[d]{Indian Institute of Science Education and Research Mohali, Knowledge City, Sector 81, SAS Nagar, Manauli 140306, India}

\emailAdd{gulab@prl.res.in}
\emailAdd{bhupal.dev@mpi-hd.mpg.de}
\emailAdd{sruba@prl.res.in}
\emailAdd{manimala@iisermohali.ac.in}

\abstract{
We analyse  in detail the  scalar triplet contribution to the low-energy 
lepton flavour violating (LFV) and lepton number violating (LNV) 
processes within a TeV-scale left-right symmetric framework.  
We show that in both type-I and type-II seesaw dominance for the light 
neutrino masses, the triplet of mass comparable to or smaller than the 
largest right-handed neutrino mass scale  can give sizeable  contribution to the LFV processes, 
except in the quasi-degenerate limit of light neutrino masses, 
where a suppression can occur due to cancellations. 
In particular, a moderate value of the heaviest neutrino to scalar triplet mass ratio $r\lesssim {\cal O}(1)$ 
is still experimentally allowed and can be explored in the future LFV  
experiments.  Similarly,  the contribution of a relatively light triplet to the LNV process of  neutrinoless double beta decay could be significant,  disfavouring a part of the model parameter space otherwise allowed by LFV constraints. 
Nevertheless, we find regions of parameter space consistent with both LFV and LNV searches, for which  
the values of the total effective neutrino mass 
can be accessible to the next generation ton-scale experiments. Such light triplets can also be directly searched for at the LHC, thus providing a complementary probe of this scenario.  Finally, we also study the implications of the triplet contribution for the left-right symmetric model interpretation of the recent diboson anomaly at the LHC. 
}

\preprint{TUM-HEP-1028/15}


\begin{document}
\maketitle

\section{Introduction}\label{sec1}

The observation of nonzero neutrino masses and mixing provides the first unambiguous experimental evidence for physics beyond the Standard Model (SM)~\cite{Agashe:2014kda}.  Although the origin of mass for all charged fermions in the SM seems to have been demystified by the Higgs boson discovery at the LHC~\cite{Aad:2012tfa,Chatrchyan:2012xdj}, the origin of tiny neutrino masses is still a nagging issue. A simple way to solve this puzzle is by breaking the global $B-L$ symmetry of the SM through Weinberg's dimension-5 operator~\cite{Weinberg:1979sa}, whose tree-level realizations are the type-I~\cite{Minkowski:1977sc, Mohapatra:1979ia, Yanagida:1979as, GellMann:1980vs, Glashow:1979nm}, II~\cite{Magg:1980ut,Schechter:1980gr,Mohapatra:1980yp,Lazarides:1980nt}  and III~\cite{Foot:1988aq} seesaw mechanisms.

A natural renormalisable theory of the effective dimension-5 operator for the seesaw mechanism is the Left-Right (L-R) Symmetric Model (LRSM) of weak interactions, based on the gauge group $SU(2)_L\times SU(2)_R\times U(1)_{B-L}$~\cite{PhysRevD.10.275,Mohapatra:1974gc,Mohapatra:1974hk,Senjanovic:1975rk}.  Here, the key ingredients of seesaw, namely, the right-handed (RH) neutrino fields, arise as the necessary parity gauge partner of the left-handed (LH) neutrino fields and are also required by anomaly cancellation, whereas the seesaw scale is identified as the one at which the  RH counterpart of the SM $SU(2)_L$ gauge symmetry, namely the $SU(2)_R$ symmetry, is broken. 
The RH neutrinos acquire a Majorana mass as soon as the $SU(2)_R$ symmetry is spontaneously broken at a scale $v_R$, analogous to the way the SM charged fermions get masses when the $SU(2)_L$ symmetry is broken at the electroweak scale $v$. Thus, the Higgs field that gives mass to the RH neutrinos becomes the analogue of the 125 GeV Higgs boson discovered at the LHC. 
 
The L-R symmetric theories lead to new effects or add new contributions to various new physics observables at both energy and intensity frontiers, which can be tested in current and future experiments, if the scale of parity restoration is below a few TeV.  In particular, a TeV-scale LRSM leads to the spectacular lepton number violating (LNV) process of same-sign dilepton plus two jets at the LHC~\cite{Keung:1983uu, Ferrari:2000sp, Nemevsek:2011hz, Chakrabortty:2012pp, Das:2012ii, AguilarSaavedra:2012gf, Chen:2013fna, Gluza:2015goa, Dev:2015kca} (for reviews, see e.g.~\cite{Senjanovic:2010nq, Deppisch:2015qwa}), as well as potentially large contributions to its low-energy analogue, namely, neutrinoless double beta decay ($0\nu\beta\beta$)~\cite{Mohapatra:1979ia, Mohapatra:1981pm, Picciotto:1982qe, Hirsch:1996qw, Tello:2010am, Chakrabortty:2012mh, Barry:2013xxa, Dev:2013vxa, Dev:2013oxa, Huang:2013kma, Dev:2014xea, Borah:2015ufa, Ge:2015yqa, Awasthi:2015ota}. In addition, there are a plethora of lepton flavour violating (LFV) processes, such as $\mu\to e\gamma$, $\mu\to 3e$, $\mu\to e$ conversion in nuclei, which can get sizeable contributions from the RH sector~\cite{Riazuddin:1981hz,
 Pal:1983bf, Mohapatra:1992uu,
 Cirigliano:2004mv, Bajc:2009ft, Tello:2010am, Chakrabortty:2012mh, Das:2012ii, Barry:2013xxa, Dev:2013oxa, Dev:2014xea, Bora:2014mna, Vasquez:2015una, Awasthi:2015ota}.  

In this paper, we focus on the scalar triplet contribution to the low-energy LNV and LFV processes within a TeV-scale LRSM framework. It is known that for triplet masses much larger than the RH neutrino masses, its contributions to LNV and LFV processes are sub-dominant~\cite{Tello:2010am,  Chakrabortty:2012mh,  Barry:2013xxa}.  However, since the direct experimental searches for these triplets 
at the LHC still allow for the possibility of low triplet masses 
$\gtrsim 500$ GeV~\cite{ATLAS:2014kca} and the current lower limits on the 
RH gauge boson masses are in the few TeV range~\cite{Khachatryan:2014dka, Aad:2015xaa, Patra:2015bga}, it is worthwhile analysing the possible scenarios 
where the triplet masses are comparable to or lower  than the 
RH neutrino or RH gauge boson masses in the theory. In such cases, we find 
that the triplet contribution to $0\nu\beta\beta$ and LFV processes can indeed 
be sizeable. While for very large RH neutrino to Higgs triplet mass ratio, these contributions are already  ruled out by existing experimental 
constraints, for moderate values of this mass ratio,  there still exists some allowed parameter space which 
can be probed in future experiments. We emphasise that these low-energy searches are  complementary to the direct probes of the scalar sector of the LRSM at colliders, where they lead to interesting multi-lepton 
signatures~\cite{Akeroyd:2005gt, Han:2007bk, Perez:2008ha, Akeroyd:2010ip, Alloul:2013raa, delAguila:2013mia, Bambhaniya:2013wza, Dutta:2014dba, Bambhaniya:2014cia, Maiezza:2015lza, Bambhaniya:2015wna}.  

The rest of the paper is  organised as follows: In Section~\ref{rev}, we review the basic features  of the minimal LRSM. In Section~\ref{lfv}, we discuss the LFV processes $\mu \to e \gamma$ and $\mu \to 3e$, and in Section~\ref{0nu2beta}, the predictions for $0\nu \beta \beta$ due to the triplet contributions. We discuss the implications for the diboson excess in Section~\ref{diboson}. Our results are summarised in Section~\ref{summary}. 

\section{The model setup} 
\label{rev}
The quarks and leptons are assigned to the following irreducible representations of the LRSM gauge group $SU(3)_c\times SU(2)_L\times SU(2)_R\times U(1)_{B-L}$~\cite{PhysRevD.10.275,Mohapatra:1974gc,Mohapatra:1974hk,Senjanovic:1975rk}: 
\begin{align}
& Q_{L,i} \ = \ \left(\begin{array}{c}u_L\\d_L \end{array}\right)_i : \: \left({ \bf 3}, {\bf 2}, {\bf 1}, \frac{1}{3}\right), \qquad \qquad  
& Q_{R,i} \ = \ \left(\begin{array}{c}u_R\\d_R \end{array}\right)_i : \: \left({ \bf 3}, {\bf 1}, {\bf 2}, \frac{1}{3}\right), \nonumber \\
& \psi_{L,i} \ = \  \left(\begin{array}{c}\nu_L \\ e_L \end{array}\right)_i : \: \left({ \bf 1}, {\bf 2}, {\bf 1}, -1 \right), \qquad \qquad 
& \psi_{R,i} \ = \ \left(\begin{array}{c} N_R \\ e_R \end{array}\right)_i : \: \left({ \bf 1}, {\bf 1}, {\bf 2}, -1 \right), 
\label{lrSM}
\end{align}
where $i=1,2,3$ represents the family index, and the subscripts $L,R$ denote  the left and right-chiral projection operators $P_{L,R} = (1\mp \gamma_5)/2$. 
For the scalar sector, the minimal model consists of the following representations:
\begin{align}
& \Phi \ = \ \left(\begin{array}{cc}\phi^0_1 & \phi^+_2\\\phi^-_1 & \phi^0_2\end{array}\right) : ({\bf 1}, {\bf 2}, {\bf 2}, 0), \nonumber \\
& \Delta_L\ = \ \left(\begin{array}{cc}\Delta^+_L/\sqrt{2} & \Delta^{++}_L\\\Delta^0_L & -\Delta^+_L/\sqrt{2}\end{array}\right) : ({\bf 1}, {\bf 3}, {\bf 1}, 2),  \nonumber  \\
& \Delta_R\ = \ \left(\begin{array}{cc}\Delta^+_R/\sqrt{2} & \Delta^{++}_R\\\Delta^0_R & -\Delta^+_R/\sqrt{2}\end{array}\right) : ({\bf 1}, {\bf 1}, {\bf 3}, 2). 
\end{align} 
The gauge symmetry $SU(2)_R\times U(1)_{B-L}$  is broken down to the SM group $U(1)_Y$ by the vacuum expectation value (VEV) of  the  neutral component of the $SU(2)_R$ triplet $\Delta_R$:   $\langle \Delta^0_R \rangle = v_R$.\footnote{In principle, the L-R symmetry can also be broken by a doublet Higgs field; however, in this case, the LH and RH neutrinos must necessarily pair up to form  Dirac particles and do not give rise to the interesting LNV signals, such as $0\nu\beta\beta$ induced by neutrinos, as discussed here. Moreover, the decay $W_R\to \ell \nu_{\ell}$ would lead to an isolated lepton plus missing energy, and the null results at the LHC in this search channel would highly disfavour a TeV-scale $W_R$ in the doublet-breaking scenario.} This  generates the Majorana masses of the RH  neutrinos $N_R$, as well as the masses of the RH gauge bosons $W_R$ and $Z_R$, and explains the small LH neutrino masses via the type-I seesaw mechanism~\cite{Minkowski:1977sc, Mohapatra:1979ia, Yanagida:1979as, GellMann:1980vs, Glashow:1979nm}.  
 The other Higgs triplet $\Delta_L$ acquires a small VEV $\langle \Delta^0_L \rangle = v_L$ and contributes to the generation of light neutrino masses via the type-II seesaw mechanism~\cite{Magg:1980ut,Schechter:1980gr,Mohapatra:1980yp,Lazarides:1980nt}. The standard electroweak symmetry is broken by the VEV of the Higgs bi-doublet field $\Phi$: $\langle\Phi\rangle={\rm diag}(\kappa_1, \kappa_2)$, which generates masses for the charged fermions, as well as the SM $W$ and $Z$ bosons. The mixing between the LH and RH gauge bosons is given by $\tan{2\xi} \simeq 2\kappa_1 \kappa_2/v_R^2$. 

The current experimental constraints on the mass of the RH gauge boson $M_{W_R}\simeq  g_Rv_R/\sqrt 2 \gtrsim 3$ TeV (assuming the equality of the $SU(2)_L$ and $SU(2)_R$ gauge couplings , i.e. $g_L=g_R$) from direct LHC searches~\cite{Khachatryan:2014dka, Aad:2015xaa}, as well as from quark flavour changing neutral current (FCNC) processes~\cite{Beall:1981ze,Zhang:2007da,Maiezza:2010ic,Bertolini:2014sua}, imply that $v_R\gtrsim 6$ TeV.  Similarly, the constraints from the electroweak $\rho$-parameter~\cite{Agashe:2014kda} restrict $v_L\lesssim 2$ GeV. On the other hand, since the VEVs of the $\Phi$ field break the electroweak symmetry, we have $\kappa_1^2+\kappa_2^2=v^2$, where $v\simeq 174$ GeV is the electroweak VEV in the SM. Thus we expect to have the following hierarchy of VEVs: 
\be
v_L \ \ll \  \kappa_{1},~\kappa_2 \ \ll \ v_R \; .
\label{eq:hie}
\ee 
Without loss of generality, we can choose $\kappa_1$ and $v_R$ as real parameters, while $\kappa_2$ and $v_L$ can, in general, be complex parameters.

The Yukawa Lagrangian in the lepton sector is given by 
\begin{eqnarray}
-{\cal L}_Y & \ = \ & h_{ij}\bar{\psi}_{L,i}\Phi \psi_{R, j} 
+ \tilde{h}_{ij}\bar{\psi}_{L, i}\tilde{\Phi} \psi_{R, j} + f_{L, ij} \psi_{L,i}^{\sf T} C i\tau_2 \Delta_L \psi_{L,j}  \nonumber \\
&& 
+ f_{R, ij} \psi_{R,i}^{\sf T} C i\tau_2 \Delta_R \psi_{R,j} 
+{\rm H.c.}, 
\label{eq:yuk}
\end{eqnarray}
where $C=i\gamma_2\gamma_0$ is the charge conjugation operator,  $\tilde{\Phi}=\tau_2\Phi^*\tau_2$, $\tau_2$ is  the  second Pauli matrix and $\gamma_{\mu}$ are the Dirac matrices. After electroweak symmetry breaking, the Yukawa Lagrangian \eqref{eq:yuk} leads to the following $6\times 6$ neutrino mass matrix in the flavour basis,
\be
{\cal M}_\nu \ = \ \left(\begin{array}{ccc}
m_L & m_D \\
m_D^{\sf T} & M_R
\end{array}\right), 
\label{eq:big}
\ee
where the $3\times 3$ Dirac and Majorana mass matrices are given by 
\be 
m_D \ = \ \frac{1}{\sqrt 2}\left(\kappa_1 h + \kappa_2 \tilde{h} \right), \quad 
m_L \ = \ \sqrt 2 v_L f_L, \quad
M_R \ = \ \sqrt 2 v_R f_R \; .
\ee
In the seesaw approximation, using  Eq.~\eqref{eq:hie}, the $3\times 3$ light neutrino 
mass matrix can be written as
\be 
m_\nu \ \simeq \ m_L - m_D M_R^{-1} m_D^{\sf T} 
 \ = \ \sqrt 2 v_L f_L - \frac{\kappa^2}{\sqrt 2 v_R} h_D f_R^{-1} h_D^{\sf T} \; , 
\label{eq:mnu}
\ee
where $h_D \equiv (\kappa_1 h+\kappa_2 \tilde{h})/(\sqrt 2 \kappa)$ and $\kappa \equiv (|\kappa_1|^2+|\kappa_2|^2)^{1/2}$. 

We will do our analysis in two interesting limits of Eq.~\eqref{eq:mnu}, which do not require any fine-tuning of the model parameters to get the observed light neutrino masses: \\[0.2pt]
{\bf (i) Type-I dominance}, where the VEV of $\Delta_L$ can be set to zero and the first term on the right-hand side of Eq.~\eqref{eq:mnu} vanishes, so that  
the light neutrino mass matrix is governed by  the  usual type-I seesaw contribution~\cite{Mohapatra:1979ia}: 
\be 
m_\nu \ \simeq \ - m_D M_R^{-1} m_D^{\sf T} \; . 
 \label{eq:type1-mnu}
\ee
In this case, the light-heavy neutrino mixing $V_{\ell N}\simeq m_DM_R^{-1}$ may or may not give large contributions to the low-energy processes, depending on the textures of $m_D$ and $M_R$ as required to satisfy the neutrino oscillation data~\cite{Dev:2014xea, Pascoli:2013fiz}.  Since our focus is on the triplet contribution, we will assume for simplicity that $m_D$ is proportional to the identity matrix~\cite{Chakrabortty:2012mh},\footnote{This could in principle be motivated from some flavour symmetry~\cite{Hagedorn:2014wha}.} with the mixing $V_{\ell N}\lesssim 10^{-6}$, which satisfies  the light neutrino mass constraint for TeV-scale $M_R$, without any fine-tuning. In this case, Eq.~\eqref{eq:type1-mnu} suggests that $m_\nu\propto M_R^{-1}$ and the same PMNS mixing matrix $U$ which diagonalises $m_\nu$ also diagonalises $M_R^{-1}$. This implies $M_R$ is diagonalised by $U^*$, since $U$ is assumed to be unitary. Moreover, the ratios of the RH neutrino mass eigenvalues ($M_i$) are related to the corresponding mass eigenvalues in the light neutrino sector ($m_i$), which are experimentally  constrained for a given mass hierarchy. Thus, the only free parameter in the RH neutrino sector is the overall mass scale, which we will 
fix by specifying the heaviest neutrino mass 
eigenvalue, to be denoted hereafter by $M_N$. More explicitly, for normal hierarchy (NH) of light neutrino masses, we have $M_N=M_1$,  and therefore, $M_2= (m_1/m_2)M_N$ and $M_3=(m_1/m_3)M_N$. Similarly, for inverted hierarchy (IH), we have $M_N=M_3$, and therefore, $M_1=(m_3/m_1)M_N$ and $M_2=(m_3/m_2)M_N$~\cite{Chakrabortty:2012mh}.
\\[0.2pt]
{\bf (ii) Type-II dominance}, when the Dirac mass term $m_D$ is negligible, so that the light neutrino mass matrix is solely governed by the Higgs triplet contribution: \be 
m_\nu \ \simeq \ m_L. 
 \label{eq:type2-mnu}
\ee
In this case, the light-heavy neutrino mixing $V_{\ell N}$ is necessarily small and does not play any role in the LNV and LFV observables. Moreover, if parity (or charge conjugation) is taken to be the discrete L-R symmetry at the TeV-scale, this implies $f_L=f_R$ (or $f_L=f_R^*$) . Hence, Eq.~\eqref{eq:type2-mnu} suggests that $m_\nu \propto M_R$, {\it i.e.},  the same PMNS mixing matrix $U$ diagonalises both LH and RH neutrino sectors~\cite{Tello:2010am}. In this case, for NH, we have $M_N=M_3$, and therefore,  $M_1= (m_1/m_3)M_N$ and $M_2=(m_2/m_3)M_N$, whereas for IH, we have $M_N=M_2$, and therefore, $M_1= (m_1/m_2)M_N$ and $M_3=(m_3/m_2)M_N$.

In the scalar sector of the minimal LRSM, there are 20 real degrees of freedom: 8 from the bi-doublet and 6 each from the LH and RH triplets. After spontaneous symmetry breaking, 6 of them are Goldstone bosons, which give masses to the LH and RH gauge bosons in both charged and neutral sectors.  Thus, there remain 14 physical real scalar fields, one of which ($h_\phi^0$) should be identified as the SM-like Higgs doublet with mass proportional to $v$, independent of the triplet VEVs. 
The remaining 13 scalar fields, {\it i.e.}, the doublets $H_\phi^0, A_\phi^0, H_\phi^\pm$, left triplets $H_L^0, A_L^0, \Delta_L^\pm, \Delta_L^{\pm\pm}$ and right triplets  $H_R^0, \Delta_R^{\pm\pm}$ are all assumed to be heavy, since their masses are proportional to $v_R$~\cite{Dev:2016dja}.  
In the following, we will be mostly interested in the masses of the doubly-charged scalars, and for simplicity, we will assume them to be equal in the LH and RH sectors. For convenience, we further define the parameter 
\begin{equation}
\frac{1}{M^2_{\Delta}} \ = \ \frac{1}{m_{\Delta_L^{\pm\pm}}^2}+\frac{1}{m^2_{\Delta_R^{\pm\pm}}} \; ,
\end{equation}
and express our results for fixed values of the ratio of the heaviest neutrino mass $M_{N}$ to $M_{\Delta}$:  $r\equiv M_N/M_\Delta$.

\section{Lepton flavour violation \label{lfv}}

\begin{figure}[t!]
\centering
\includegraphics[width=12cm]{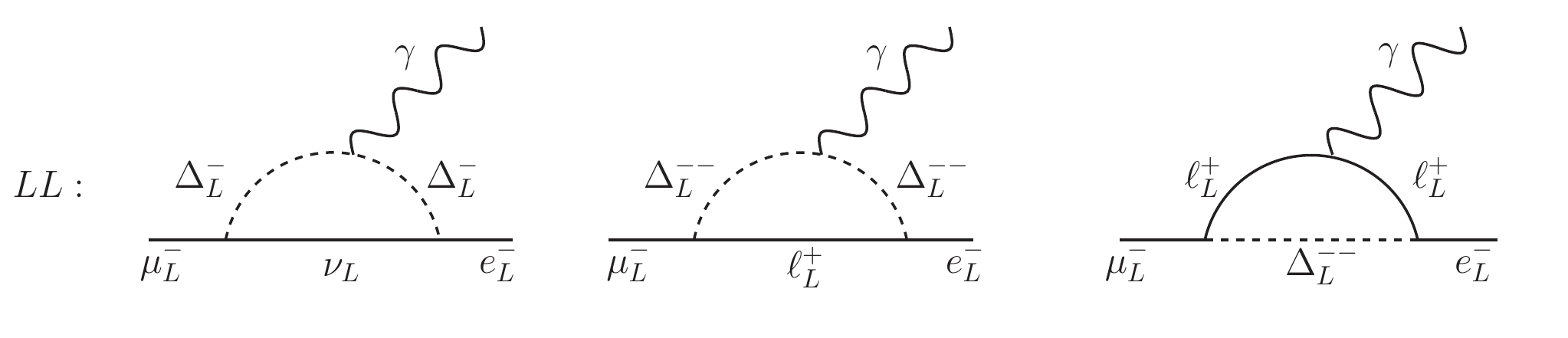}\\
\includegraphics[width=12cm]{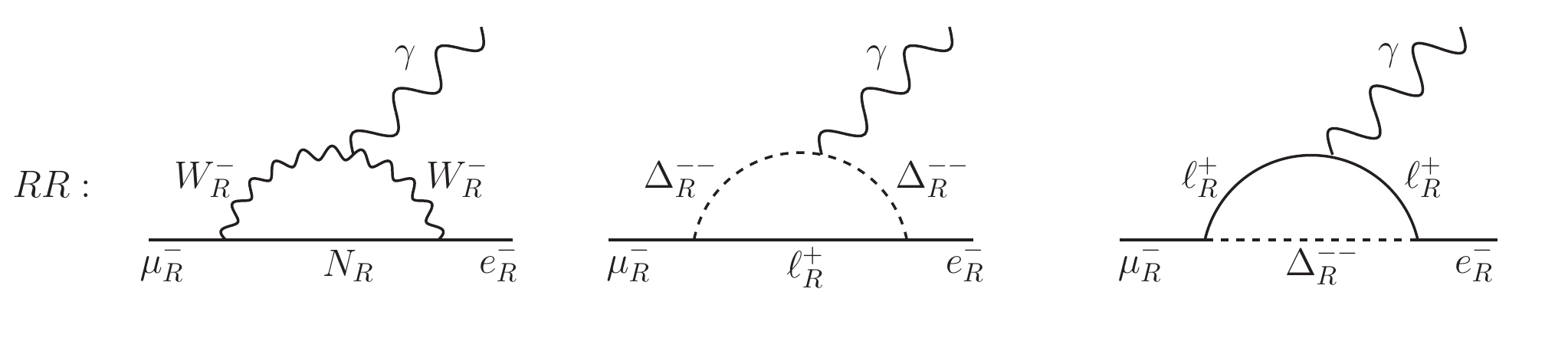}\\
\includegraphics[width=5cm]{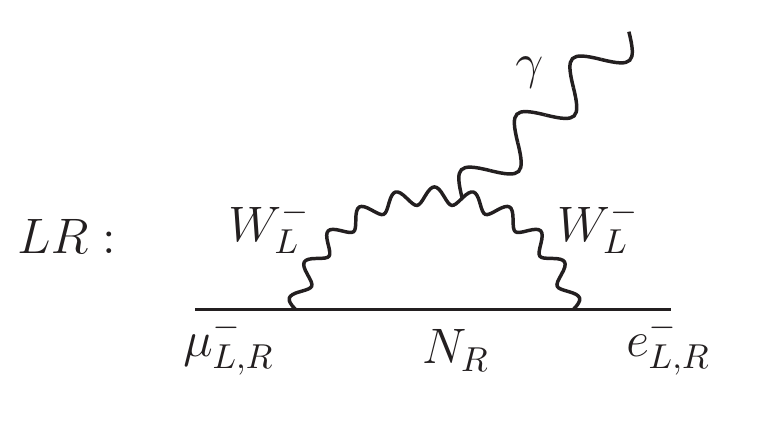}
\caption{Feynman diagrams for $\mu\to e\gamma$ in the LRSM.}
\label{fig:1}
\end{figure}

In the canonical SM seesaw, the LFV decay rates induced by the neutrino mixing are suppressed by the tiny neutrino masses, and hence, are well below the current experimental limits \cite{Bellgardt:1987du,Adam:2013mnn} and even the distant-future sensitivities \cite{Baldini:2013ke,Kuno:2005mm,Blondel:2013ia}. On the other hand, in the LRSM, several new contributions 
   appear due to the additional RH current interactions, which could lead to sizeable LFV rates for a TeV-scale $v_R$. For example, the $\mu\to e\gamma$ process receives new contributions from both the scalar and gauge sectors, which can be classified into three categories, namely, those involving purely LH currents ($LL$), purely RH currents ($RR$) and mixed LH-RH currents ($LR$), as shown in Figure~\ref{fig:1}. The corresponding branching ratio is given by~\cite{Cirigliano:2004mv, Barry:2013xxa}
\begin{equation}
 \textrm{BR}({\mu \to e \gamma}) \ = \ 
\frac{3\alpha_{\rm em}}{2\pi} \left(  \abs{G_L^{\gamma}}^2 + \abs{G_R^{\gamma}}^2  \right) ,
\label{eq:mutoegamma}
\end{equation}
where $\alpha_{\rm em}\equiv e^2/4\pi$ is the electromagnetic coupling constant, and the form factors $G^{\gamma}_R$ and $G^{\gamma}_L$ are given by 
\begin{align}
 G_R^{\gamma} & \ =    \sum_{i=1}^{3} \bigg({V}_{\mu i} V_{e i}^*|\xi^2|G_1^{\gamma}(a_i) - S^*_{\mu i}{V}_{e i}^* \xi e^{-i \alpha} G_2^{\gamma}(a_i)\frac{M_i}{m_{\mu}} 
+ {V}_{\mu i} V^*_{e i} \bigg[\frac{m^2_{W_L}}{m^2_{W_R}}  G_1^{\gamma}(b_i) + \frac{2b_i}{3} \frac{m_{W_L}^2}{m_{\Delta_R^{++}}^2}\bigg]   \bigg), \label{grgamma} \\
G_L^{\gamma} & \  =    \sum_{i=1}^{3} \bigg({S}^*_{\mu i} {S}_{e i}G_1^{\gamma}(a_i) - {V}_{\mu i} S_{e i} \xi e^{i \alpha} G_2^{\gamma}(a_i)\frac{M_i}{m_{\mu}} 
+ {V}_{\mu i} V^*_{e i} b_i \bigg[\frac{2}{3} \frac{m^2_{W_L}}{m^2_{\Delta_L^{++}}}  + \frac{1}{12} \frac{m_{W_L}^2}{m_{\Delta_L^{+}}^2}\bigg]   \bigg), \label{glgamma}
\end{align}
with $a_i \equiv (M_i/m_{W_L})^2$, $b_i \equiv (M_i/m_{W_R})^2$, $\alpha$ is the phase of the VEV  $\kappa_2$, $m_\mu$ is the muon mass, $V$ is the RH neutrino mixing matrix which is related to the PMNS mixing matrix in our case, and $S$ is the light-heavy neutrino mixing matrix which can be neglected for the choice of our parameters. Similarly, we can drop the terms depending on the $W_L-W_R$ mixing parameter $\xi$ which is experimentally constrained to be $\lesssim 10^{-3}$~\cite{Agashe:2014kda}. 
The loop functions $G_{1,2}^{\gamma}(a)$ are given as
\begin{eqnarray}
 G_1^{\gamma}(a) & \ = \ & -\frac{2a^3+5a^2-a}{4(1-a)^3} - \frac{3a^3}{2(1-a)^4} ~\textrm{ln} ~a \;, \\
G_2^{\gamma}(a) & \ = \ & \frac{a^2-11a+4}{2(1-a)^2} - \frac{3a^2}{(1-a)^3 } ~\textrm{ln} ~a \; .
\label{loopfucnt}
\end{eqnarray}

\begin{figure}[t!]
\centering
\includegraphics[width=12cm]{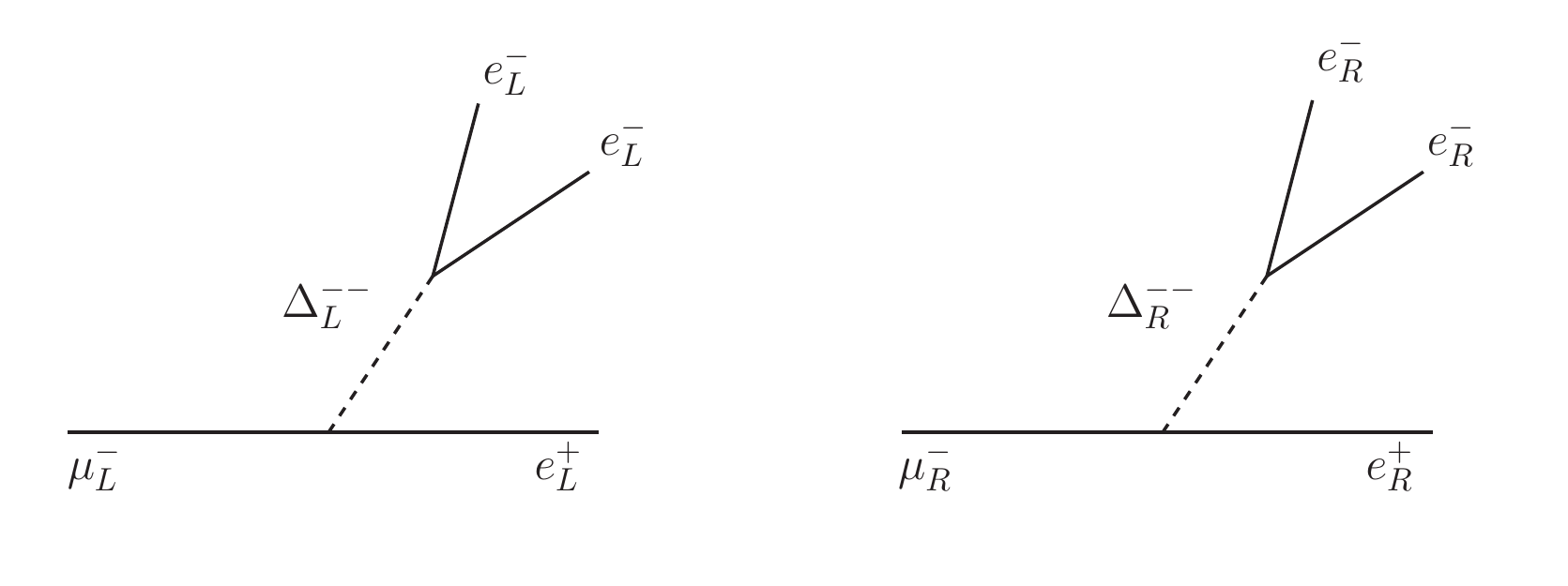}
\caption{Feynman diagrams for $\mu\to 3e$ in the LRSM.}
\label{fig:2}
\end{figure}

For the LFV process $\mu \to 3 e$, the Higgs triplets $\Delta_L$ and $\Delta_R$  
contribute
at the tree level, as shown in Figure~\ref{fig:2}, thereby making the branching ratio of this process potentially large~\cite{Cirigliano:2004mv,Leontaris:1985qc, Swartz:1989qz, Cirigliano:2004tc}: 
\begin{eqnarray}
\textrm{BR}(\mu \to 3e) \ = \ \frac{1}{2}\left|h_{\mu e}h^*_{ee}\right|^2\left(\frac{m^4_{W_L}}{m^4_{\Delta^{++}_L}}+\frac{m^4_{W_L}}{m^4_{\Delta^{++}_R}}\right) \; ,
\label{eq:mu3e}
\end{eqnarray}
where $h_{\alpha \beta}\equiv \sum_{i=1}^3V_{\alpha i}V_{\beta i}M_i/m_{W_R}$. Note that there is also an one-loop induced contribution in the type-I dominance~\cite{Ilakovac:1994kj}, which is however suppressed by the loop factors as well as by the light-heavy neutrino mixing, and hence, we can safely ignore it in our case, as compared to the tree-level contribution given by Eq.~\eqref{eq:mu3e}.  
In  Ref.~\cite{Tello:2010am}, it 
has been pointed out that the current experimental constraint on ${\rm BR}(\mu\to 3e)\leq 1.0\times 10^{-12}$~\cite{Bellgardt:1987du} requires that in Eq.~\eqref{eq:mu3e}, the triplet scalar masses must be at least 10 times the heaviest RH neutrino mass scale in the theory, {\it i.e.}, the ratio $r\lesssim 0.1$, thereby 
making the Higgs triplet contribution to $\mu \to e \gamma$ and  $0 \nu \beta \beta$ negligible. We show that while this is true in general, there can be cancellations due to the variations of the so far unknown $C\!P$ phases in the PMNS mixing matrix in which cases, this is not strictly required, {\it i.e.}, the $\mu\to 3e$ rate can in principle be compatible with the experimental constraint even for larger values of $r$. 
In these interesting scenarios, the Higgs triplet contribution to other LFV and  $0\nu \beta \beta$ processes  can become sizeable, and hence, must be included in the analysis.   
This is first illustrated with three representative values of $r$ (moderate, small and large), where we show that 
$r$ values as large as ${\cal O}(1)$ are still allowed by current experimental constraints, giving rise to interesting effects in low-energy LNV and LFV observables, as well as potential LNV signals at the LHC. Then we show the LFV-allowed parameter space as a function of the ratio $r$. We do not explicitly discuss here other interesting LFV processes, such as $\mu-e$ conversion in nuclei, or electric dipole moments, which are left for future studies. 

\label{sec:moderate_r}
\begin{figure}[t!]
\subfloat[]{\includegraphics[width=7.1cm, angle =0]{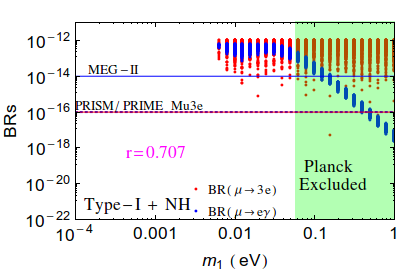}\label{021a}} 
\subfloat[]{\includegraphics[width=7.1cm, angle =0]{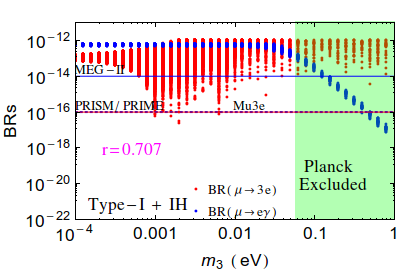}\label{021b}} \\
\subfloat[]{\includegraphics[width=7.1cm, angle =0]{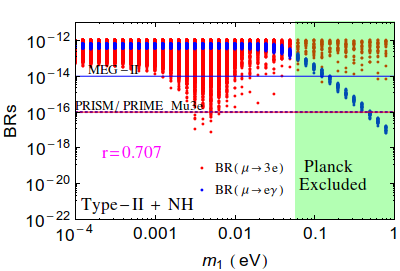}\label{021c}} 
\subfloat[]{\includegraphics[width=7.1cm, angle =0]{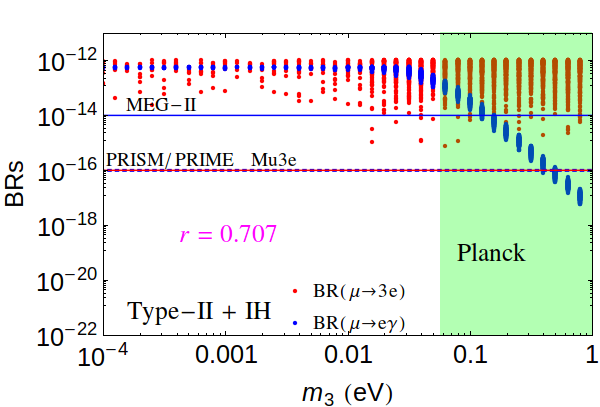}\label{021d}} 
\caption{The predicted branching ratios of $\mu \to e\gamma$ (blue points) and $\mu \to 3e$ (red points)  processes (when for a given light neutrino mass, current experimental bounds on the branching ratios of both are simultaneously satisfied) as a function of the lightest neutrino mass for NH (left panels) and IH (right panels) in type-I (top panels) and type-II (bottom panels) dominance. The ratio of the heaviest neutrino mass and the Higgs triplet mass has been set to  $r=0.707$. The green shaded region is disfavoured at 95\% C.L. from Planck data. The blue solid horizontal line is for MEG-II sensitivity, while PRISM/PRIME and Mu3e will have sensitivities up to the blue dotted and red solid horizontal lines respectively.} 
\label{fig:lfvmodr}
\end{figure}

\subsection*{Case-I: Moderate value of ${r}$} 
We first consider the scenario with $r=0.707$. For illustration, we set the 
RH gauge boson mass $m_{W_R}=3.5$ TeV,  largest heavy neutrino mass $M_{N}=500$ GeV and the Higgs triplet masses $M_{\rm scalar}\equiv m_{\Delta_R^{++}}=m_{\Delta_L^{++}} = M_{\Delta_L^{+}} = 1$ TeV, which are consistent with the direct experimental constraints from the LHC. Using these parameters and Eqs.~\eqref{eq:mutoegamma} and \eqref{eq:mu3e}, we compute the $\mu\to e\gamma$ and $\mu\to 3e$ branching ratios, respectively, as a function of the lightest neutrino mass. We have taken into account the $3\sigma$ variation of the oscillation parameters as given by a recent global fit~\cite{Forero:2014bxa}, as well as the variation of the Dirac $C\!P$ phase $\delta$ between [0, $2\pi$]
and Majorana phases $\alpha_{1,2}$ between [0, $\pi$]. We demand that our predicted LFV branching ratios  should satisfy the current limits: ${\rm BR}(\mu\to e\gamma)<5.7\times 10^{-13}$ from MEG~\cite{Adam:2013mnn} and ${\rm BR}(\mu\to 3e)<1.0\times 10^{-12}$ from SINDRUM~\cite{Bellgardt:1987du} experiments. Our results are shown in Figure~\ref{fig:lfvmodr} by the blue ($\mu\to e\gamma$) and red ($\mu\to 3e$) scattered points for normal hierarchy (NH, left panels) and inverted hierarchy (IH, right panels) in type-I (top panels) and type-II (bottom panels) dominance. 
We find that for the type-I, NH case, the predicted LFV branching ratios of $\mu \to e \gamma$ and $\mu \to 3e$ are allowed by  the present experimental constraints, only if the  lightest neutrino mass $m_1 \ge 0.01$ eV. For all other cases, lower values of $m_1$($m_3$) are allowed. A part of this parameter space with quasi-degenerate neutrinos is disfavoured by the most stringent limit on the sum of light neutrino masses $\Sigma_i m_i < 0.17$ eV at 95$\%$ C.L from 
Planck data~\cite{Ade:2015xua}, as shown by the green shaded region in Figure~\ref{fig:lfvmodr}. 
We infer that for moderate values of $r$, the predicted LFV  branching ratios for both type-I and type-II dominance are within the reach of future experiments, such as MEG-II~\cite{Baldini:2013ke}, PRISM/PRIME~\cite{Kuno:2005mm} and Mu3e~\cite{Blondel:2013ia}, as shown by the blue and red horizontal lines in Figure~\ref{fig:lfvmodr}.

\begin{figure}[t!]
\begin{center}
\subfloat[]{\includegraphics[width=7.5cm, angle =0]{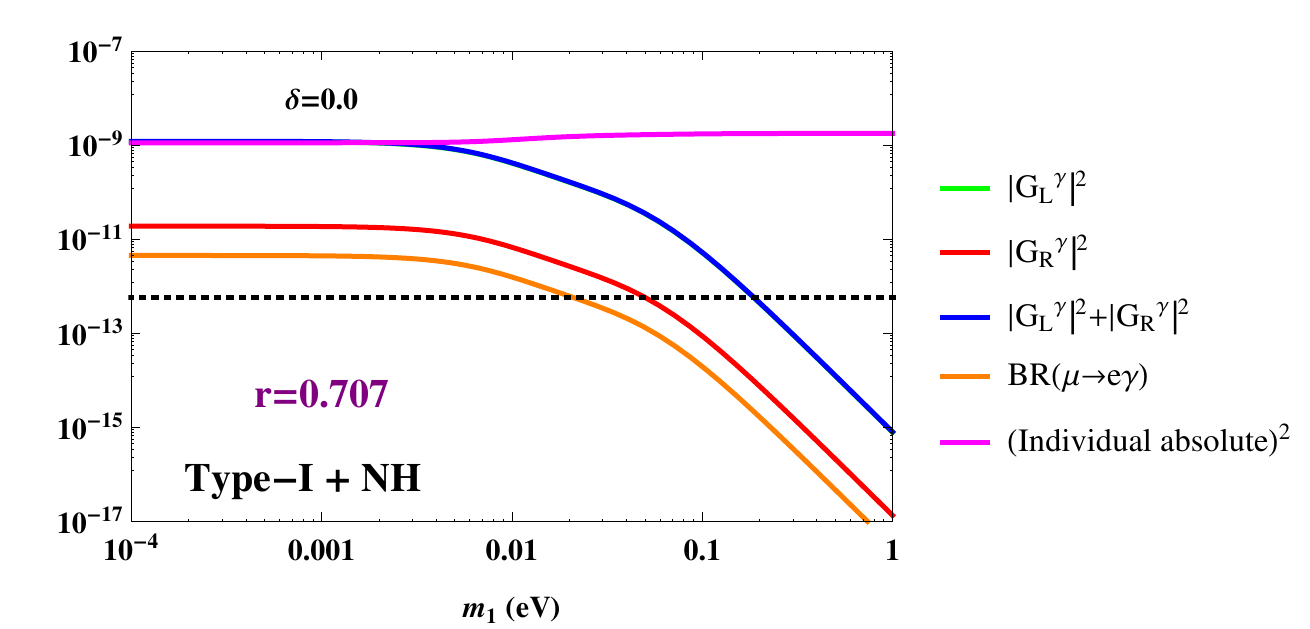} \label{fig3p}} 
\subfloat[]{\includegraphics[width=7.5cm, angle =0]{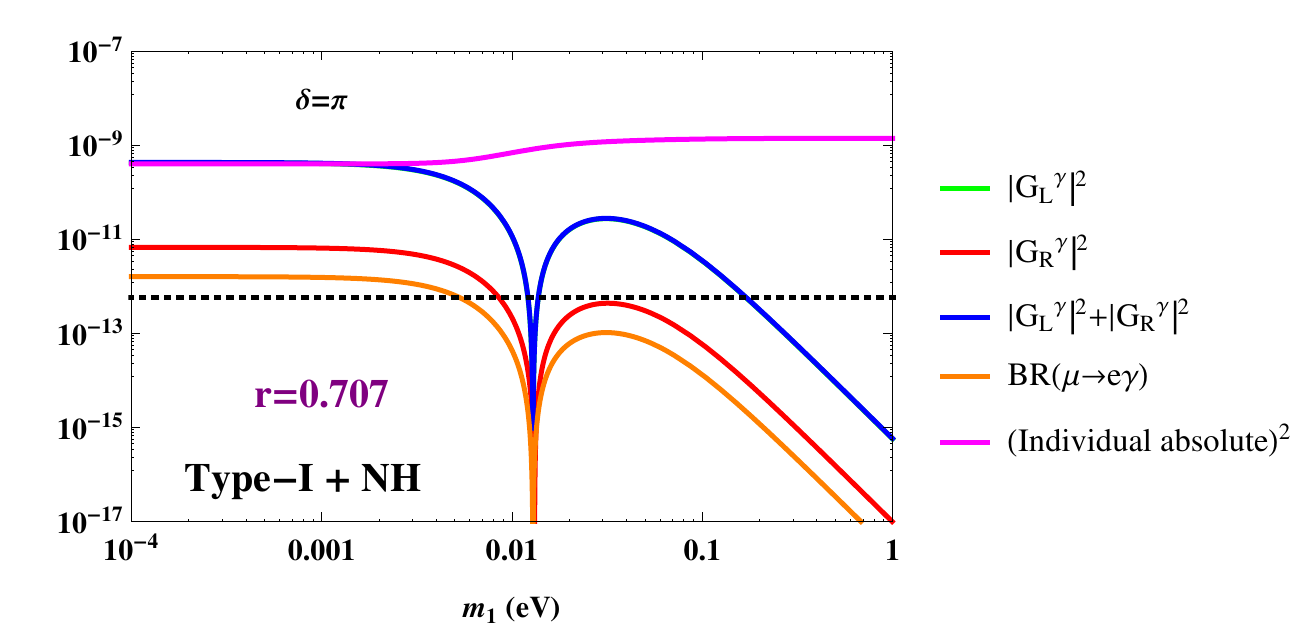} \label{fig3q}}\\
\subfloat[]{\includegraphics[width=6cm, angle =0]{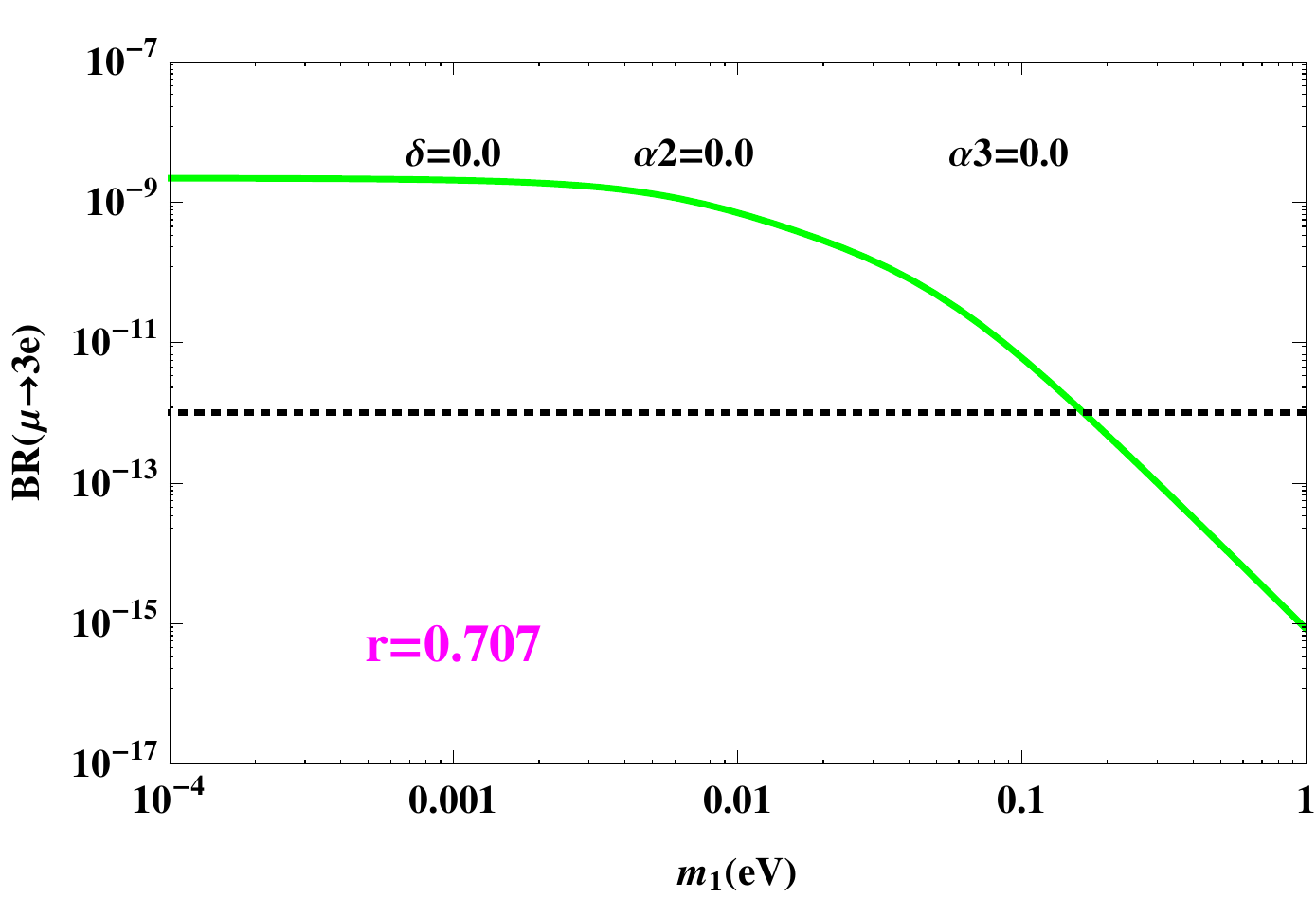} \label{figr}}\hspace{1cm}
\subfloat[]{\includegraphics[width=6cm, angle =0]{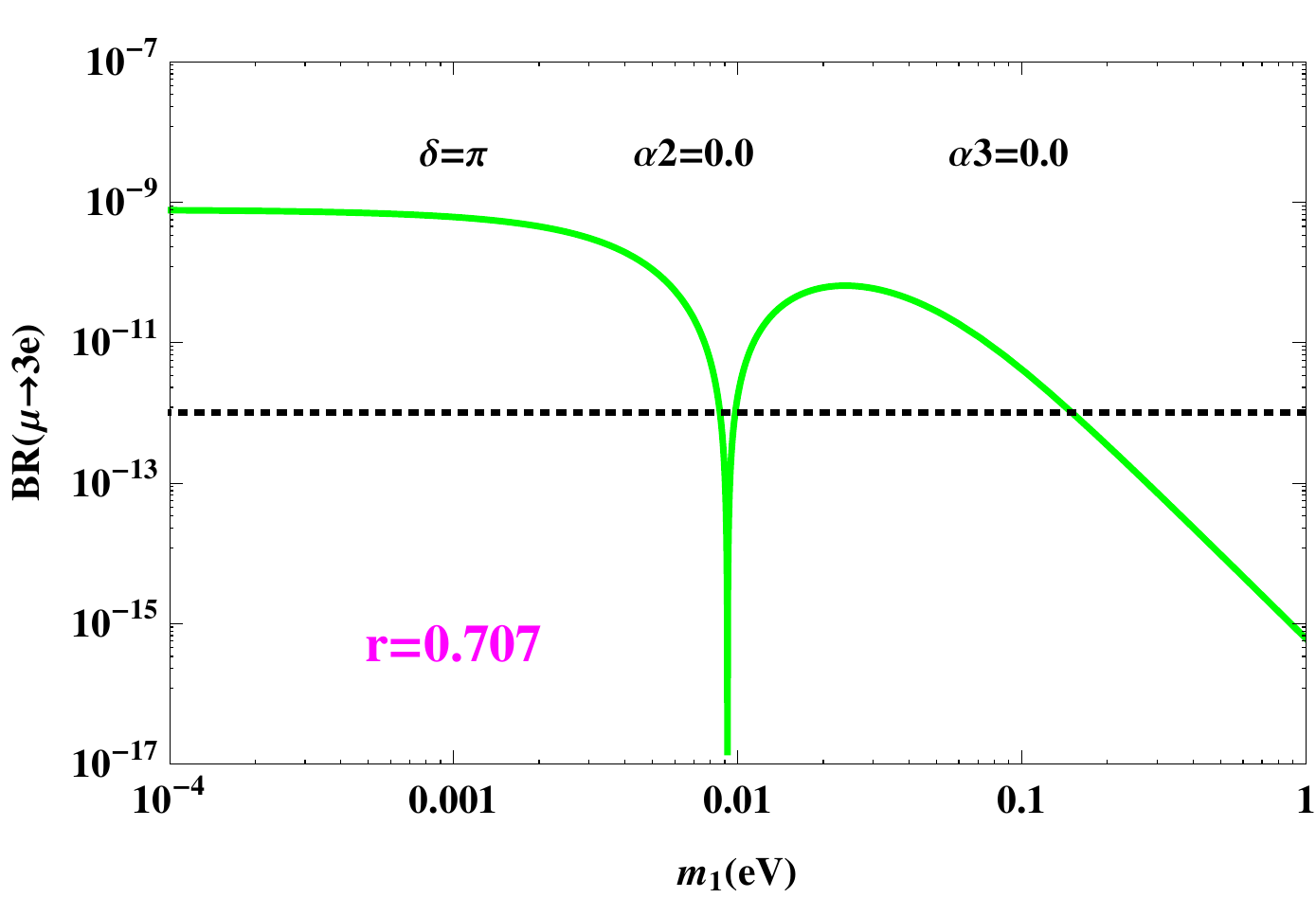} \label{figt}}
\end{center}
\caption{{\em Upper panels:} Variation of $G^{\gamma}_L$, $G^{\gamma}_R$ and the total branching ratio of $\mu \to e \gamma$ process as a function of the lightest neutrino mass for $\delta=0$ (left) and $\pi$ (right).  {\em Lower panels:} Variation of the branching ratio of $\mu \to 3 e $ as a function of the lightest neutrino mass for $\delta=0$ (left) and $\pi$ (right). Here we have chosen $\alpha_2 = 0$, $\alpha_3 = 0$, $r = 0.707$, type-I dominance and  NH case. 
}
 \label{fig:abc}  
\end{figure}

To better understand the dependence of the branching ratios on the lightest neutrino mass, next we consider only the best-fit values of the oscillation 
parameters, as depicted in Figures~\ref{fig3p} and \ref{fig3q}, where we show the  individual contributions $G^{\gamma}_L$, $G^{\gamma}_R$ [{\it cf.} Eqs.~\eqref{grgamma} and \eqref{glgamma}] to the  branching ratio of $\mu \to e \gamma$, as well as the total contribution, for two different $C\!P$ violating 
phases. For the line labeled as  $\rm{(individual~absolute)}^2$, we have 
summed over the absolute-square of the individual contributions 
 inside  $G^{\gamma}_L$, $G^{\gamma}_R$, thereby  neglecting the possibility of any interference.
However, the interference terms are indeed important for the total contribution to the LFV 
branching ratio, as can be seen from Figure~\ref{fig:abc}. The phase variation  induces  a suppression in the  branching ratio due to  cancellation between different contributions.  We highlight this particular 
feature with suitable choices  of the  $C\!P$ phases $\delta=0$ and  $\pi$ in Figures~\ref{fig3p} and \ref{fig3q}, from which it is evident that,  while the (individual absolute)$^2$ increases with the lightest neutrino mass, the contributions $G^{\gamma}_L$, $G^{\gamma}_R$ 
as well as the total  $\rm{BR}(\mu \to e \gamma)$ decrease for quasi-degenerate light neutrino masses. 
Similar feature is visible  for $\mu \to 3e$ process, as depicted in Figures~\ref{figr} and \ref{figt}. From Figures~\ref{fig3q} and \ref{figt}, it is evident that  for the Dirac $C\!P$ phase $\delta=\pi$, there is an additional suppression in the branching ratios of $\mu \to e \gamma$ and $\mu \to 3e$ near $m_1 \sim 0.01 ~\rm{eV}$ due to an exact cancellation between  different terms.

\subsection*{Case-II: Smaller value of   $r$}
\begin{figure}[t!]
\subfloat[]{\includegraphics[width=7.1cm, angle =0]{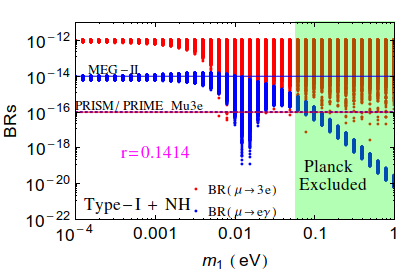}\label{11a}}
\subfloat[]{\includegraphics[width=7.1cm, angle =0]{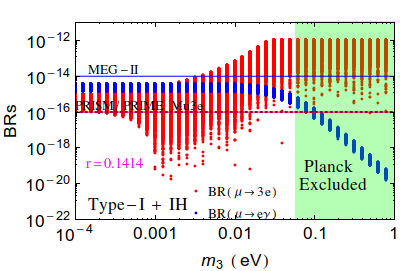}\label{11b}}\\
\subfloat[]{\includegraphics[width=7.1cm, angle =0]{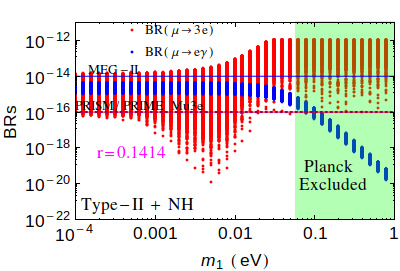}\label{11c}}
\subfloat[]{\includegraphics[width=7.1cm, angle =0]{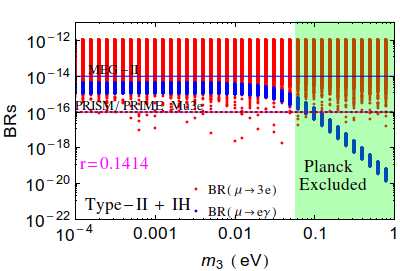}\label{11d}}
\caption{The predicted branching ratios of $\mu \to e\gamma$ (blue points) and $\mu \to 3e$ (red points)  processes as a function of the lightest neutrino mass for NH (left panels) and IH (right panels) in type-I (top panels) and type-II (bottom panels) dominance. The ratio of the heaviest neutrino mass and the Higgs triplet mass has been set to  $r=0.1414$. The green shaded region is disfavoured at 95\% C.L. from Planck data. 
The blue solid horizontal line is for MEG-II sensitivity, while PRISM/PRIME and Mu3e will have sensitivities up to the blue dotted and red solid horizontal lines respectively.  }
\label{fig:lfvlowr}
\end{figure}

Next we consider the case where 
 $M_N = 500~\textrm{GeV}$ and $M_{\rm scalar}= 5~\textrm{TeV}$, leading to $r=0.1414$. For such a heavy  Higgs triplet, we expect its contribution to LFV processes to be relatively smaller, thereby allowing more LRSM parameter space for hierarchical neutrinos. This is indeed the case, as shown in  Figure~\ref{fig:lfvlowr}.  
A few comments are in order: 
(i) For the process $\mu \to e \gamma$, the predicted branching ratio is beyond the reach of MEG-II upgrade~\cite{Baldini:2013ke} excepting for type-I dominance and  NH [{\it cf.} Figure~\ref{11a}], where hierarchical $m_1$ ($\lsim 0.01 ~ \rm{eV}$)  may just
be within its reach. However, for the process $\mu \to 3e$, the predicted branching ratios are within the experimental reach 
of  Mu3e~\cite{Blondel:2013ia}. 
(ii) For the scenarios shown in Figures~\ref{11a}--\ref{11c}, an additional suppression occurs due to phase cancellation 
 in the branching ratio of $\mu \to 3e$  for for  $m_{\rm{lightest}} \sim 10^{-3}-10^{-2}$ eV, thereby making part of the allowed parameter space  beyond the reach of 
the Mu3e sensitivity. However,  the type-II dominance IH cases is  not affected by such phase-cancellation [{\it cf.} Figure~\ref{11d}], and hence, can be tested more easily in future. 

\begin{figure}[t!]
\subfloat[]{\includegraphics[width=7.1cm, angle =0]{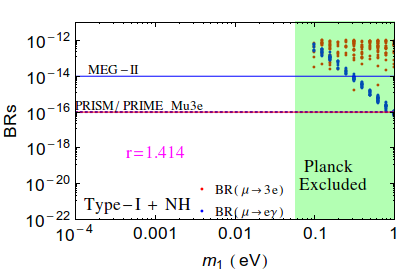}\label{17a}}
\subfloat[]{\includegraphics[width=7.1cm, angle =0]{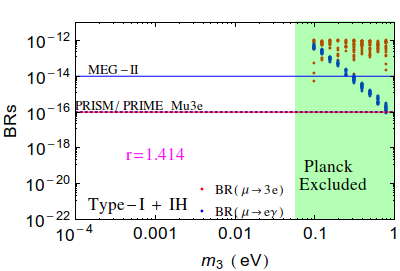}\label{17b}}\\
\subfloat[]{\includegraphics[width=7.1cm, angle =0]{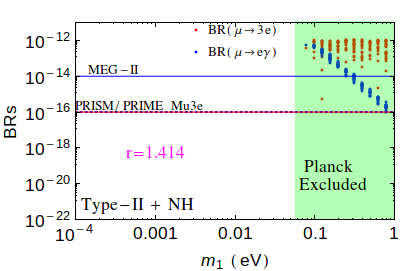}\label{17c}}
\subfloat[]{\includegraphics[width=7.1cm, angle =0]{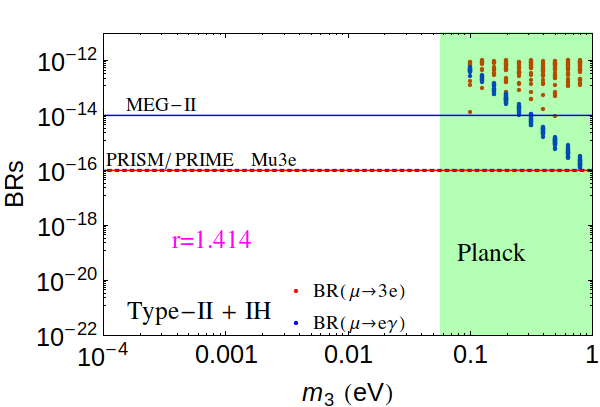}\label{17d}}
\caption{The predicted branching ratios of $\mu \to e\gamma$ (blue points) and $\mu \to 3e$ (red points)  processes as a function of the lightest neutrino mass for NH (left panels) and IH (right panels) in type-I (top panels) and type-II (bottom panels) dominance. The ratio of the heaviest neutrino mass and the Higgs triplet mass has been set to  $r=1.414$. The green shaded region is disfavoured at 95\% C.L. from Planck data. 
The blue solid horizontal line is for MEG-II sensitivity, while PRISM/PRIME and Mu3e will have sensitivities up to the blue dotted and red solid horizontal lines respectively.}
\label{fig:lfvhighr}
\end{figure}

\subsection*{Case-III: Larger value of $r$}

In Figure~\ref{fig:lfvhighr}, we show the prediction for the other interesting regime, {\it i.e.}, lighter Higgs triplet and heavier RH neutrinos. We consider $M_N =500~\textrm{GeV}$ and 
$M_{\rm scalar}=500~ \textrm{GeV}$, so that $r=1.414$. In this case, the predicted LFV rates will be much larger than the previous two cases, due to a large triplet contribution. Hence, this scenario is heavily constrained from present experimental constraints. It is evident from Figure~\ref{fig:lfvhighr} that the 
predicted branching ratios are in agreement with the experimental LFV rates, only for quasi-degenerate mass regime, which is already disfavoured by the cosmological constraints from Planck.

\begin{figure}[t!]
\subfloat[]{\includegraphics[width=7.1cm, angle =0]{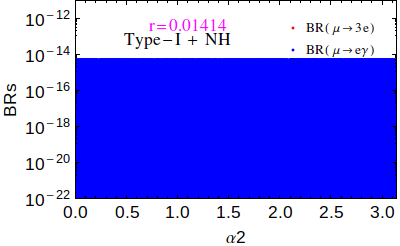}\label{17aaa}}
\subfloat[]{\includegraphics[width=7.1cm, angle =0]{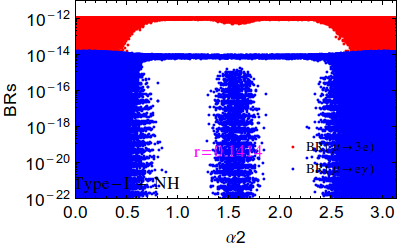}\label{17bbb}}\\
\subfloat[]{\includegraphics[width=7.1cm, angle =0]{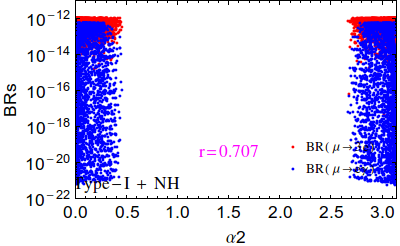}\label{17ccc}}
\subfloat[]{\includegraphics[width=7.1cm, angle =0]{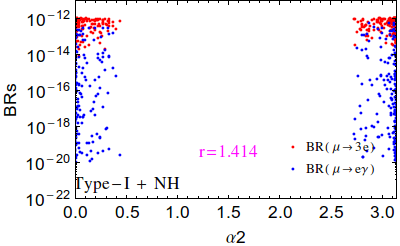}\label{17ddd}}
\caption{The predicted branching ratios of $\mu \to e\gamma$ (blue points) and $\mu \to 3e$ (red points) processes, when experimental bounds on the branching ratios of both are simultaneously satisfied, as a function of the Majorana phase $\alpha_2$ for type-I NH case and with different values of $r$.}
 \label{fig:alpha2}  
\end{figure}

Depending on the value of $r$, one can also obtain some constraints on the Majorana phase $\alpha_2$ from the LFV bounds, as illustrated in Figure~\ref{fig:alpha2} for the type-I NH case with $M_N=500$ GeV. The oscillation parameters are varied as before and $m_1$ is varied in the range $10^{-4}$ eV to 1 eV. Figure~\ref{17aaa} shows that 
for $r=0.01414$, corresponding to $M_{\rm scalar} = 50 ~\rm{TeV}$,
there are no constraints from LFV processes as for such a heavy mass, the triplet is effectively decoupled. As the value of $r$ increases the allowed values of $\alpha_2$ start getting restricted from LFV constraints
and the preferred values for $\alpha_2$  are seen to cluster around 
0 and $\pi$. For $r=1.414$, the LFV constraints are stronger and the density of the points is lesser. We did not find any such constraints on the phase $\alpha_1$ from the LFV muon decays. 

\begin{figure}[t!]
\subfloat[]{\includegraphics[width=7.1cm, angle =0]{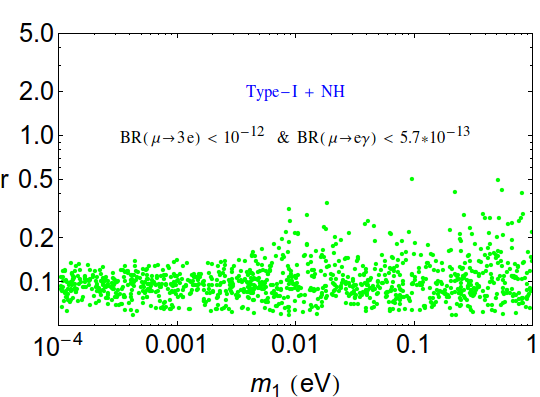}\label{rma}}
\subfloat[]{\includegraphics[width=7.1cm, angle =0]{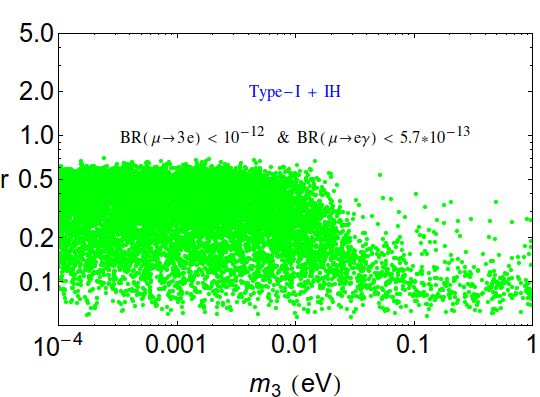}\label{rmb}}\\
\subfloat[]{\includegraphics[width=7.1cm, angle =0]{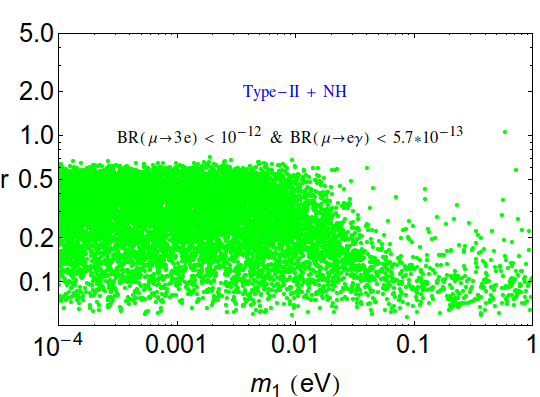}\label{rmc}}
\subfloat[]{\includegraphics[width=7.1cm, angle =0]{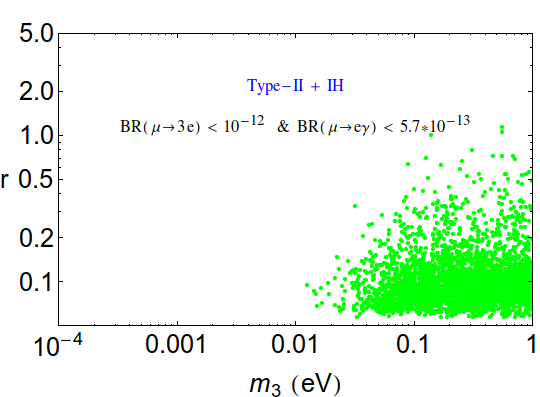}\label{rmd}}
\caption{The allowed parameter space as a function of $r$ satisfying both $\mu \to e\gamma$ and $\mu \to 3e$ constraints simultaneously.}
 \label{fig:rmsmall}  
\end{figure}

To summarize our findings in this section, values of $r$ up to ${\cal O}(1)$  can be still allowed by the LFV constraints, depending on other parameters in the light neutrino mass matrix. This is illustrated with respect to the variation in $r$ in Figure~\ref{fig:rmsmall}, where the scattered points simultaneously satisfy both $\mu\to e\gamma$ and $\mu\to 3e$ constraints for $M_N=500$ GeV.  

\section{Neutrinoless double beta decay \label{0nu2beta}}
\begin{figure}[t!]
\centering
\subfloat[]{\includegraphics[width=4.1cm, angle =0]{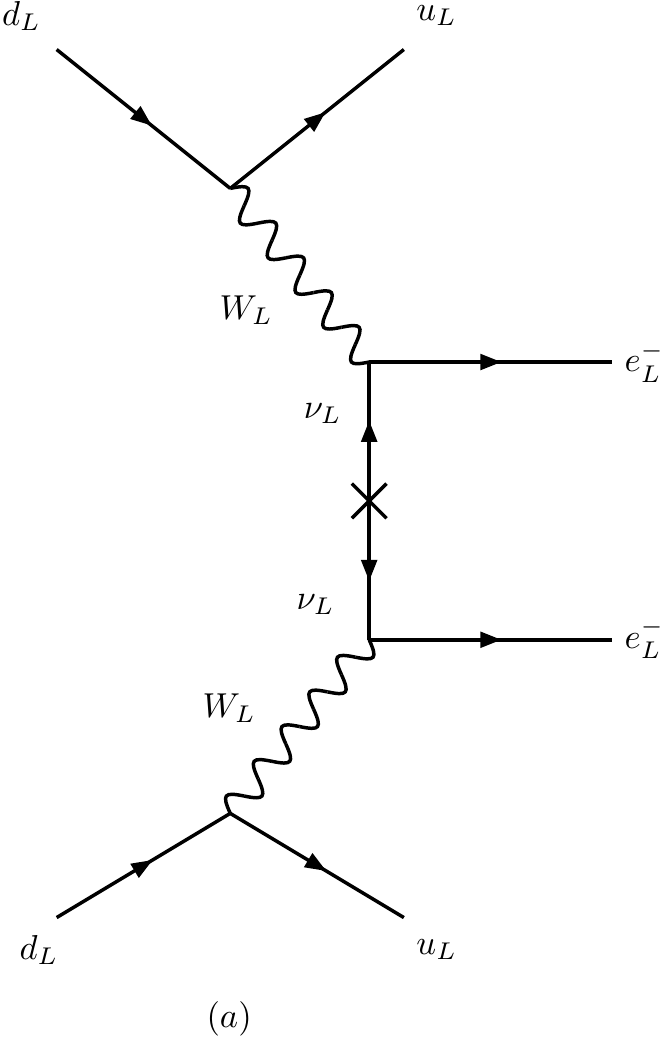}\label{0nu2betaa}}
\hspace{0.5cm}
\subfloat[]{\includegraphics[width=4.1cm, angle =0]{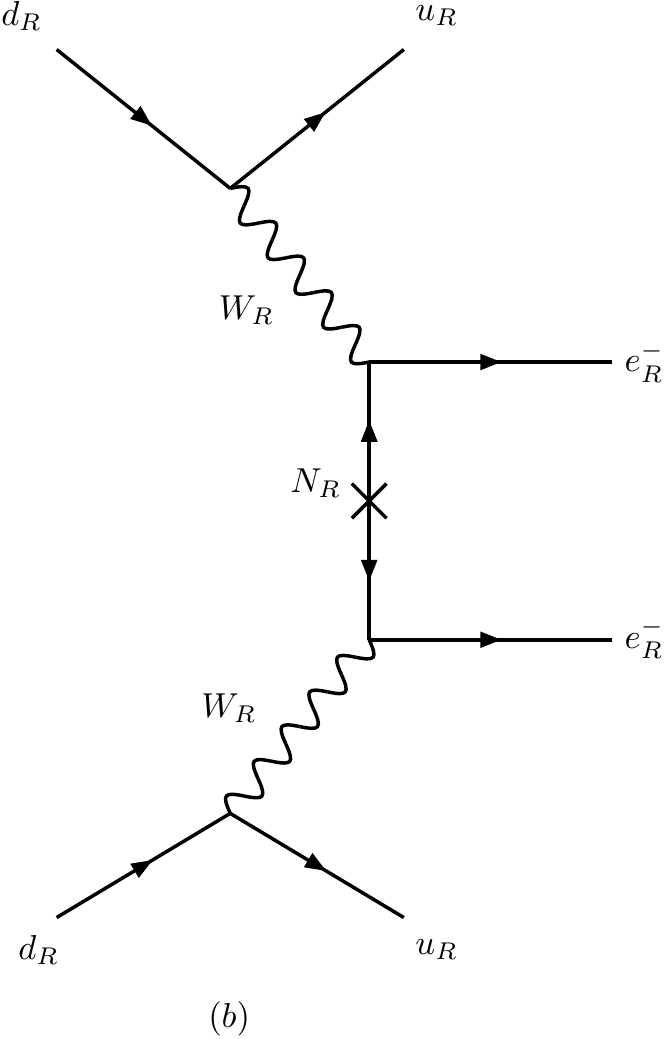}\label{0nu2betab}} 
\subfloat[]{\includegraphics[width=5.1cm, angle =0]{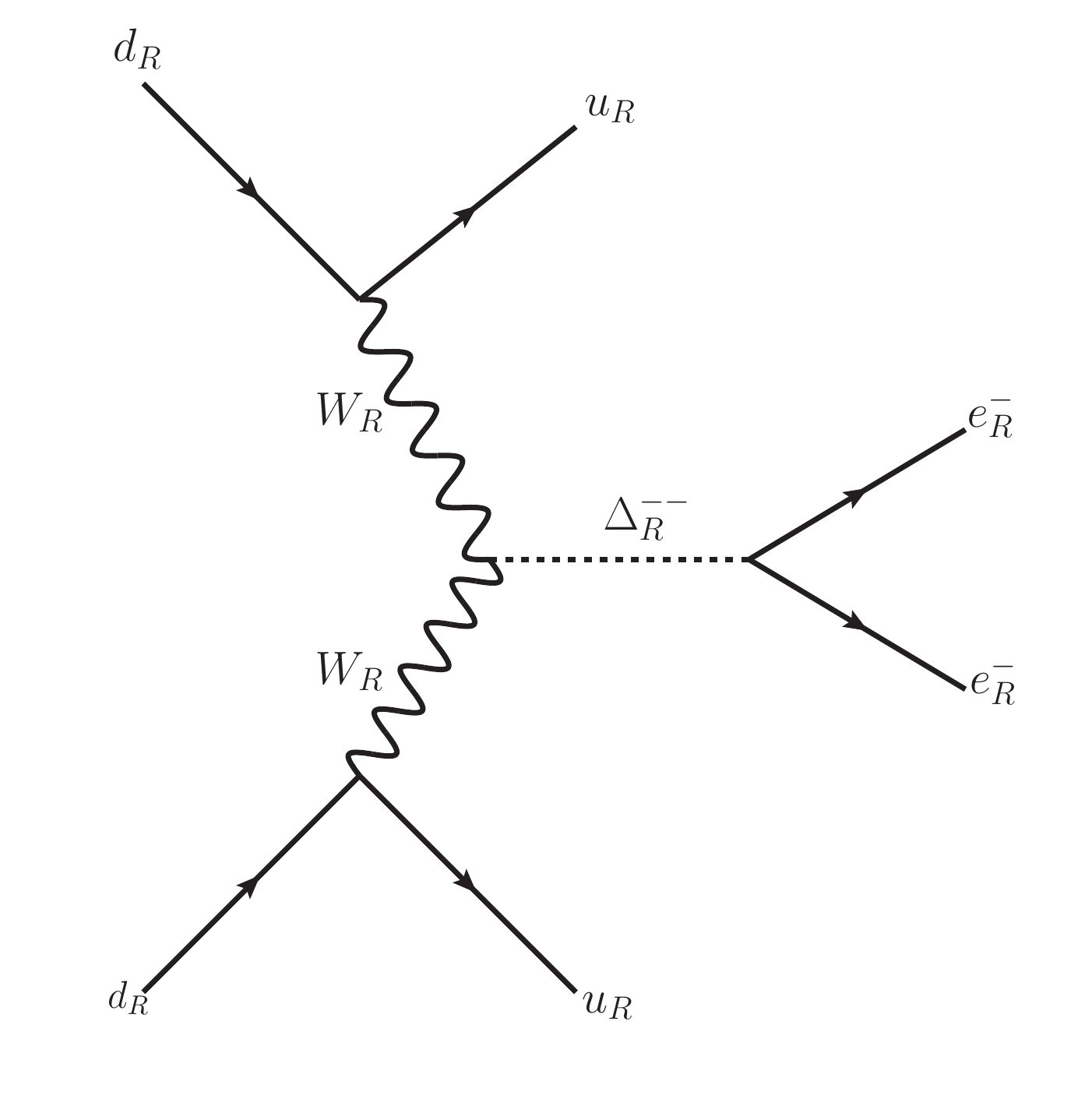}\label{0nu2betad}} 
\caption{The dominant LH and RH current contributions to the $0\nu \beta \beta$ process in the LRSM with small LH-RH mixing. }
\label{fig:diagrams2}
\end{figure}

In a TeV-scale LRSM, there are  several new contributions to the LNV process of $0\nu \beta \beta$~\cite{Mohapatra:1979ia,Mohapatra:1981pm,Picciotto:1982qe, Hirsch:1996qw, Tello:2010am, Chakrabortty:2012mh, Barry:2013xxa, Dev:2013vxa, Dev:2013oxa, Huang:2013kma, Dev:2014xea, Borah:2015ufa, Ge:2015yqa, Awasthi:2015ota}, due to the presence of RH  currents and  Higgs triplets. 
\begin{figure}[t!]
\subfloat[]{\includegraphics[width=7.1cm, angle =0]{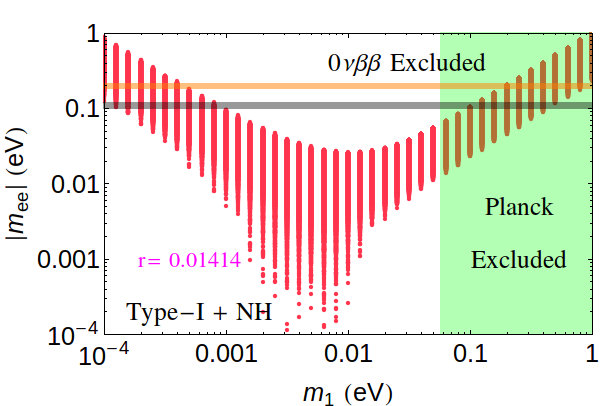}\label{21a}}
\subfloat[]{\includegraphics[width=7.1cm, angle =0]{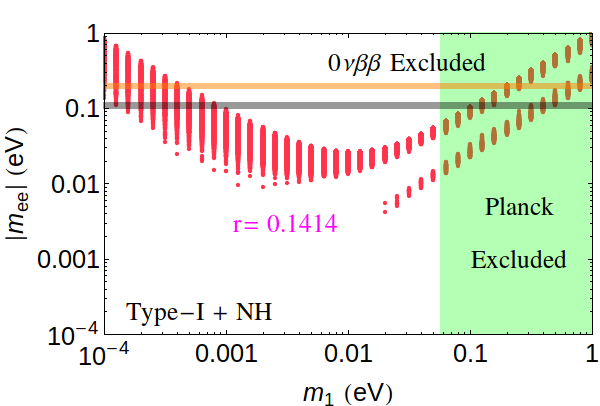}\label{21b}}\\
\subfloat[]{\includegraphics[width=7.1cm, angle =0]{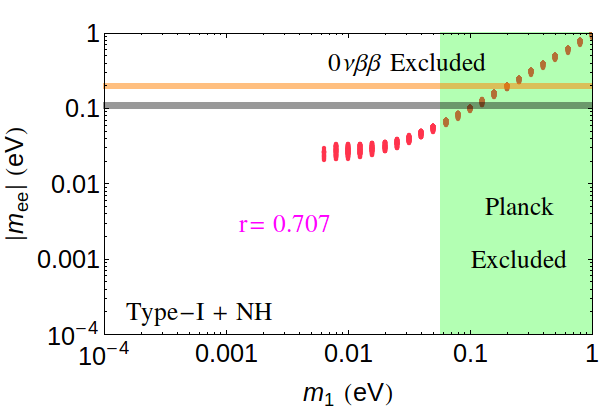}\label{21c}} 
\subfloat[]{\includegraphics[width=7.1cm, angle =0]{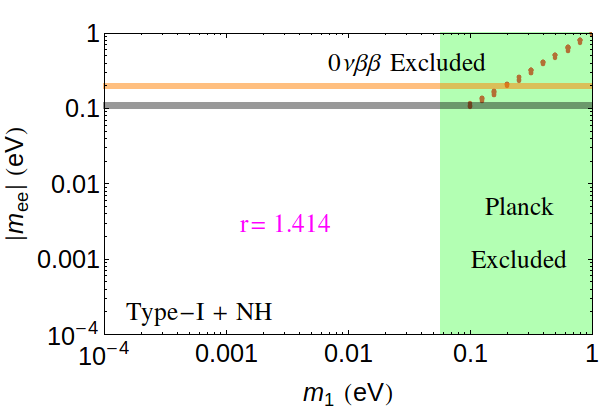}\label{21d}}
\caption{The variation of the effective neutrino mass as a function of the lightest neutrino mass  for  type-I dominance with NH. The different panels correspond to different values of $r$. The green shaded area is disfavoured at 95\% C.L. by Planck. The orange band corresponds to the range of $|m_{ee}| = 0.18 - 0.22$ eV, the region above which is excluded at 90\% C.L. by 
the combined limit from GERDA. The black band corresponds to  the future limit ($|m_{ee}| = 0.098 - 0.12$ eV) from GERDA-II.
} 
\label{fig:meet1nh}
\end{figure}
As discussed in the previous section, the present bounds from $\mu\to e\gamma$ and $\mu \to 3e$ still allow the heavy neutrino to  Higgs triplet masses   
as large as ${\cal O}(1)$. So the Higgs triplet contribution to $0\nu \beta \beta$ can in principle be sizeable and should not be neglected. In our subsequent discussion of $0 \nu \beta \beta$, we therefore take into account the Higgs triplet contribution from $\Delta_R$. The 
contribution from the other Higgs triplet $\Delta_L$ is suppressed by the light neutrino mass. Also we assume the mixing between the LH and RH sectors to be small, so that their contributions to $0 \nu \beta \beta$ can be neglected.

Thus, in our case, the half-life 
of $0\nu \beta \beta$ only includes purely LH and RH contributions: 
\be
\frac{1}{T_{1/2}^{0\nu}} \ = \ G_{01}^{0\nu} \Bigg(\Big|{\cal M}_\nu^{0\nu}\eta_\nu \big|^2 + \Big|{\cal M}_N^{0\nu}\eta_{R}\Big|^2   \Bigg) ,
\label{half}
\ee   
where $G_{01}^{0\nu}$ is the phase space factor and ${\cal M}^{0\nu}_{\nu,N}$ are the relevant nuclear matrix elements (NMEs) for light and  heavy neutrino contributions, respectively. The particle physics parameters  $\eta_{\nu}$ and $\eta_R$ correspond to the 
LH and RH amplitudes, respectively ({\it cf.} Figure~\ref{fig:diagrams2}):  
\begin{eqnarray}
\eta_\nu  \ = \ \frac{1}{m_e}\sum_{i=1}^3 U_{ei}^2 m_i,  \qquad  
\eta_{R} \ = \  m_p \left(\frac{m_{W_L}}{m_{W_R}}\right)^4 \bigg ( \sum_{i=1}^3  \frac{V_{ei}^2}{M_i}+ \sum_{i=1}^3 \frac{V_{ei}^2 M_i} {m^2_{\Delta_R^{++}}} \bigg),
\label{etanu}
\end{eqnarray}
where $m_e$ and $m_p$ are the masses of electron and proton, respectively. 
The corresponding effective neutrino mass is given by 
\begin{eqnarray}
m_{ee} \ = \ \sum_i U_{ei}^2 m_i+ \langle p^2 \rangle  \left(\frac{m_{W_L}}{m_{W_R}}\right)^4 \bigg ( \sum_i  \frac{V_{ei}^2}{M_i}+ \sum_i\frac{V_{ei}^2 M_i} {m^2_{\Delta_R^{++}}} \bigg) ,
\label{mee}
\end{eqnarray}
where $\langle p^2 \rangle = m_e m_p {\cal M}_N^{0\nu}/{\cal M}_\nu^{0\nu} \sim (153-184~\rm{MeV})^2$ for $^{76}$Ge isotope~\cite{Meroni:2012qf} which we have taken as our reference nucleus in this analysis. 
 
In Figure~\ref{fig:meet1nh}, we show the effective mass $m_{ee}$ versus the 
lightest neutrino mass $m_1$ for type-I dominance with NH 
and for different values of the ratio $r$. In this and all other 
figures in this section for obtaining the effective mass, 
we have used only those values of the model parameters    
that are consistent with the 
experimental limits of $\mu \to e\gamma$ and $\mu \to 3e$ processes, as discussed in Section~\ref{lfv}. 
Thus these plots are inclusive of the LFV constraints. We have included the $3\sigma$ variation of the oscillation parameters from Ref.~\cite{Forero:2014bxa}, as well as the NME uncertainties as reported in Ref.~\cite{Meroni:2012qf}.

\begin{figure}[t!]
\subfloat[]{\includegraphics[width=7.1cm, angle =0]{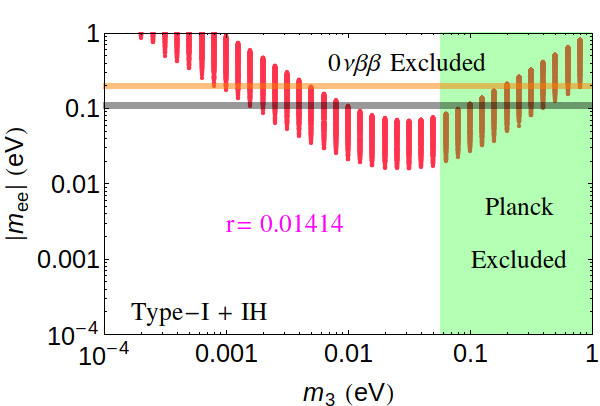}\label{12d}}
\subfloat[]{\includegraphics[width=7.1cm, angle =0]{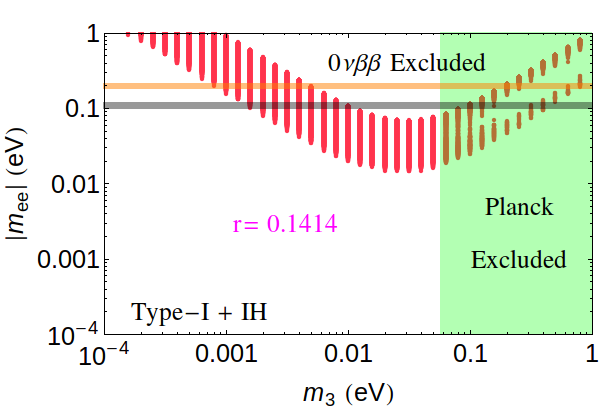}\label{12a}}\\
\subfloat[]{\includegraphics[width=7.1cm, angle =0]{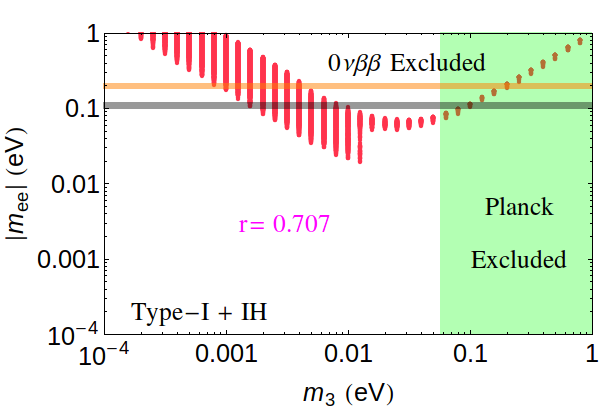}\label{12b}} 
\subfloat[]{\includegraphics[width=7.1cm, angle =0]{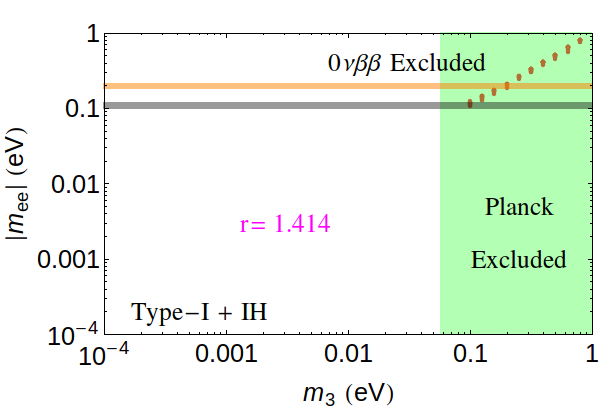}\label{12c}}
\caption[The variation of the effective mass as a function of the lightest neutrino mass for  type-I dominance and IH.]{ The variation of the effective mass as a function of the lightest neutrino mass for  type-I dominance and IH.  The different panels correspond to different values of $r$. The green shaded area is disfavoured at 95\% C.L. by Planck. The orange band corresponds to the range of $|m_{ee}| = 0.18 - 0.22$ eV, the region above which is excluded at 90\% C.L. by the combined limit from GERDA.
 The black band corresponds to  the future limit ($|m_{ee}| = 0.098 - 0.12$ eV) from GERDA-II. The bands are due to the NME uncertainties. }
\label{fig:meet1ih}
\end{figure}

\begin{figure}[t!]
\subfloat[]{\includegraphics[width=7.1cm, angle =0]{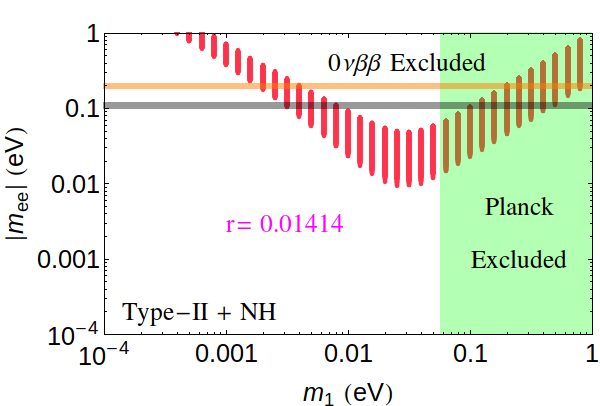}\label{13d}}
\subfloat[]{\includegraphics[width=7.1cm, angle =0]{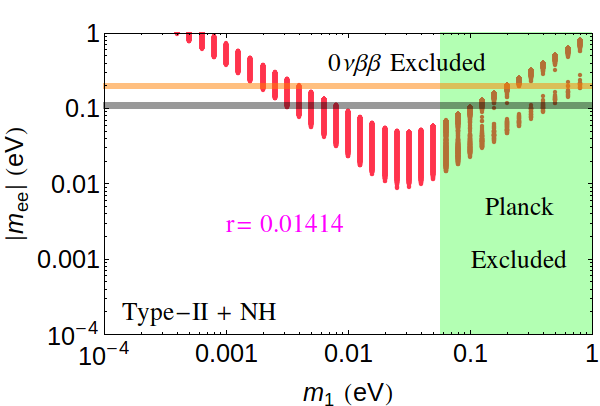}\label{13a}}\\
\subfloat[]{\includegraphics[width=7.1cm, angle =0]{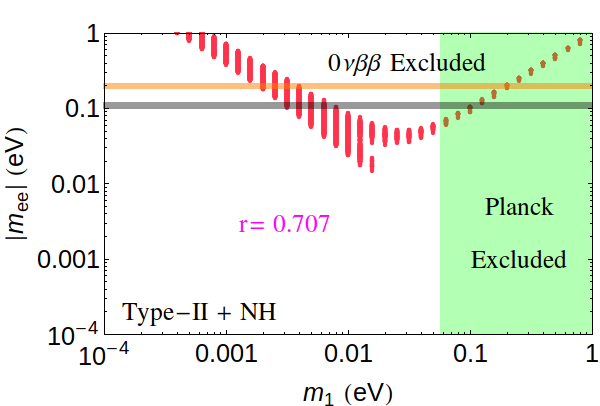}\label{13b}} 
\subfloat[]{\includegraphics[width=7.1cm, angle =0]{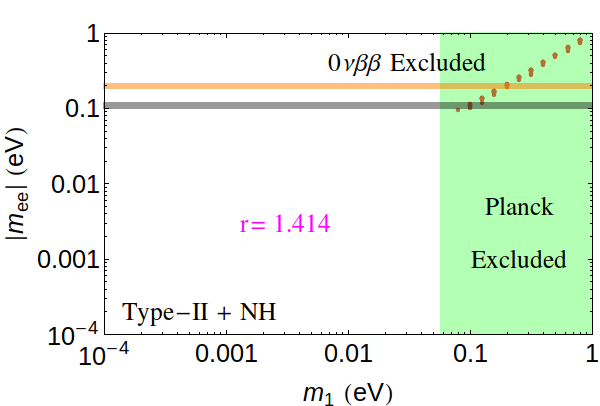}\label{13c}}
\caption[The variation of the effective mass as a function of the light neutrino mass for type-II dominance and NH.]{ The variation of the effective mass as a function of the light neutrino mass for type-II dominance and NH. The orange band corresponds to the range of $|m_{ee}| = 0.18 - 0.22$ eV, the region above which is excluded at 90\% C.L. by 
the combined limit from GERDA.
 The black band corresponds to  the future limit ($|m_{ee}| = 0.098 - 0.12$ eV) from GERDA-II. }
\label{fig:meet2nh}
\end{figure}

Figure~\ref{21a} is for $r= 0.01414$ ($M_N= 500 ~ \rm{GeV}$, $M_{\rm scalar} = 50 ~\rm{TeV}$). Such a heavy  triplet is almost decoupled, and hence, there are no additional constraints on $0\nu\beta\beta$ 
from the LFV processes. 
Thus in this case, the effective mass $m_{ee}$  is the same as that obtained in Refs.~\cite{Chakrabortty:2012mh, Dev:2014xea} without including the triplet contribution. Note however that, although  there  are no constraints from LFV processes, the current $0\nu\beta\beta$ bounds from GERDA~\cite{Agostini:2013mzu} disfavour lower (fully hierarchical) and higher (quasi-degenerate) values of $m_1$. The quasi-degenerate region is also disfavoured from Planck data. The future limits from GERDA-II~\cite{Majorovits:2015vka} could even place a stronger {\em lower} limit on the lightest neutrino mass in this scenario. 

\begin{figure}[t!]
\subfloat[]{\includegraphics[width=7.1cm, angle =0]{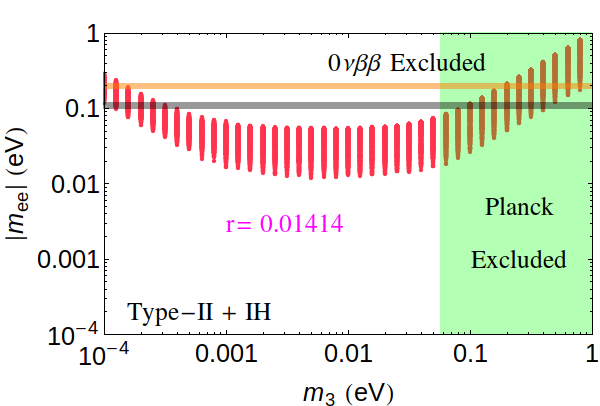}\label{14d}}
\subfloat[]{\includegraphics[width=7.1cm, angle =0]{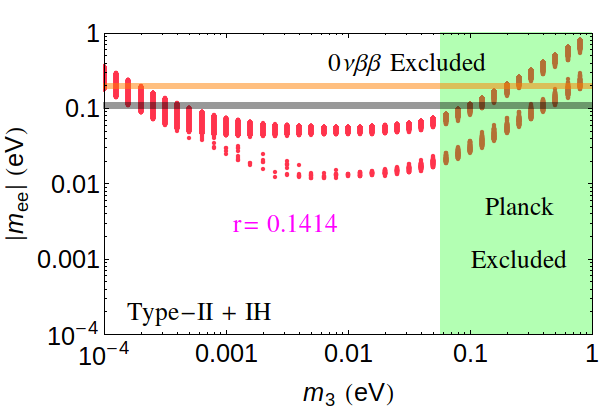}\label{14a}}\\
\subfloat[]{\includegraphics[width=7.1cm, angle =0]{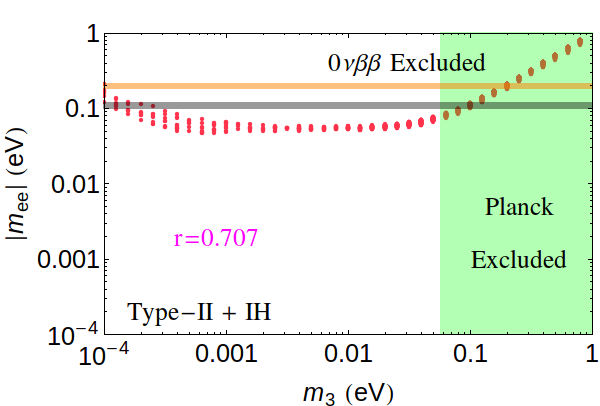}\label{14b}} 
\subfloat[]{\includegraphics[width=7.1cm, angle =0]{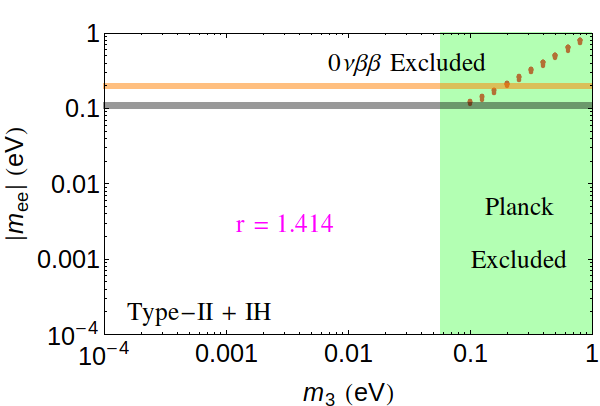}\label{14c}}
\caption[The variation of the effective mass as a function of the lightest neutrino mass for  type-II  dominance and IH.]{ The variation of the effective mass as a function of the lightest neutrino mass for  type-II  dominance and IH.  The different panels correspond to different values of $r$. The green shaded area is disfavoured at 95\% C.L. by Planck. The orange band corresponds to the range of $|m_{ee}| = 0.18 - 0.22$ eV, the region above which is excluded at 90\% C.L. by 
the combined limit from GERDA.
 The black band corresponds to  the future limit ($|m_{ee}| = 0.098 - 0.12$ eV) from GERDA-II. }
\label{fig:meet2ih}
\end{figure}

As we go to a higher value of $r= 0.1414$ ($M_N= 500 ~ \rm{GeV}$, $M_{\rm scalar} = 5 ~\rm{TeV}$), as shown in Figure~\ref{21b}, the current LFV constraints (see  Figure~\ref{fig:lfvlowr}) still allow the whole range of $m_1$. 
However,  there are additional constraints on the 
Majorana phase $\alpha_2$ as has been shown in Figure~\ref{fig:alpha2}.  
This rules out a part of the parameter space involving the cancellation region, and therefore, very low values 
of $m_{ee}$ can no longer be  obtained.  The shape of the curve for $|m_{ee}|$  in Figure~\ref{21b} 
can be solely attributed to the LFV constraints on the Majorana phases.
We have checked that if LFV constraints are not included, then Figure~\ref{21b}   replicates Figure~\ref{21a}.

For smaller Higgs triplet masses that lead  to  
larger value of $r$, such as, $r=$ 0.707 and 1.1414, 
the hierarchical mass range $m_1 \le 0.01 ~\rm{eV}$  is 
completely ruled out and only 
the quasi-degenerate region  is allowed by the LFV constraints, as shown in the first panel corresponding to 
type-I NH in Figure~\ref{fig:lfvmodr} and  Figure~\ref{fig:lfvhighr}. 
The corresponding impact of the LFV constraints on the prediction for $0\nu \beta \beta$ 
is clearly visible from  Figure~\ref{21c} and Figure~\ref{21d}, 
where the effective mass is in agreement with the LFV constraints mostly for quasi-degenerate  light neutrino masses. 
Note that most of this region is already disfavoured by the Planck data and/or the current upper limit on $m_{ee}$ from GERDA. 
For r=0.707 a small window for $m_1$  ($\sim 0.005 - 0.05$ eV) 
still exists which is consistent with all the current constraints.
However this region is beyond the reach of GERDA-II and might be accessible only with future ton-scale experiments, such as MAJORANA+GERDA~\cite{Abgrall:2013rze}.

Similarly, in Figures~\ref{fig:meet1ih},  \ref{fig:meet2nh}  and  \ref{fig:meet2ih}, 
 we show the effective mass versus lightest neutrino mass for the case of type-I dominance with IH, type-II dominance with NH and IH, respectively.
In all these scenarios,  the $r=0.01414$ case again resembles to the cases where the Higgs triplet effect is not included~\cite{Chakrabortty:2012mh, Dev:2014xea}. Also note that 
for these plots the cancellation region with very low value of $m_{ee}$ 
is not obtained.  The exclusion of certain regions of parameter 
space specially for higher values of the lightest neutrino mass is due to the 
constraint on the phase $\alpha_2$ from LFV processes, as explicitly shown in Figure~\ref{fig:alpha2}. 

From Figures~\ref{fig:meet1nh}, \ref{fig:meet1ih},  \ref{fig:meet2nh}  and  \ref{fig:meet2ih}, it is evident that a large value of $r$ is highly constrained experimentally, whereas a moderate value of $r\lesssim {\cal O}(1)$ is more favourable and 
can  be tested in the next generation $0\nu \beta \beta$ experiments, such as GERDA-II~\cite{Majorovits:2015vka}, in combination with the future LFV experiments. 

\begin{figure}[t!]
\includegraphics[width=7.1cm, angle =0]{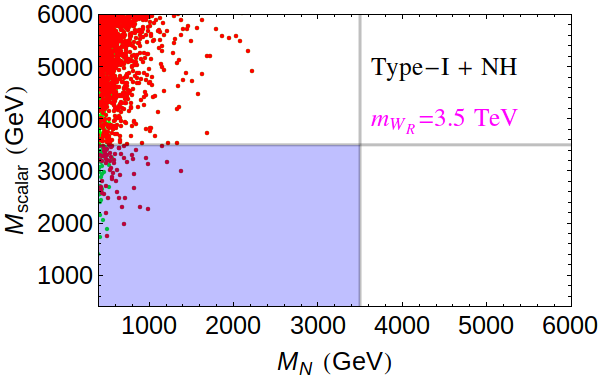}
\includegraphics[width=7.1cm, angle =0]{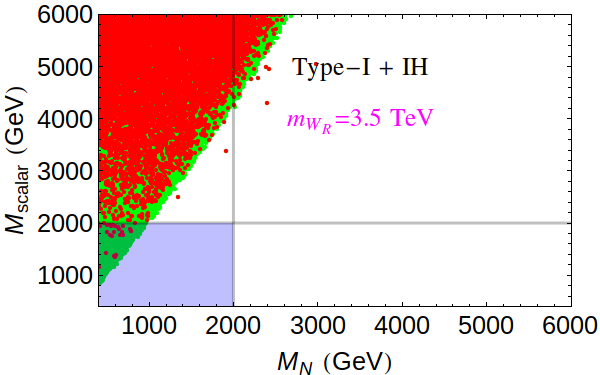}\\
\includegraphics[width=7.1cm, angle =0]{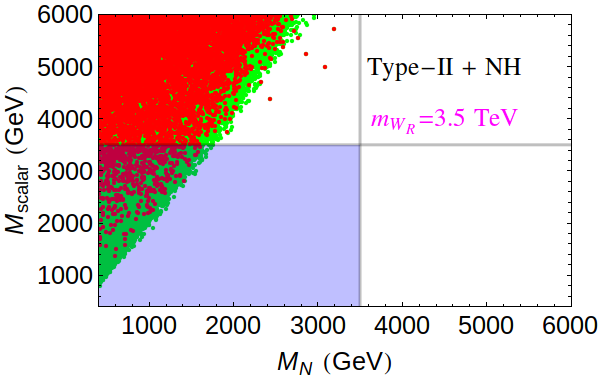}
\includegraphics[width=7.1cm, angle =0]{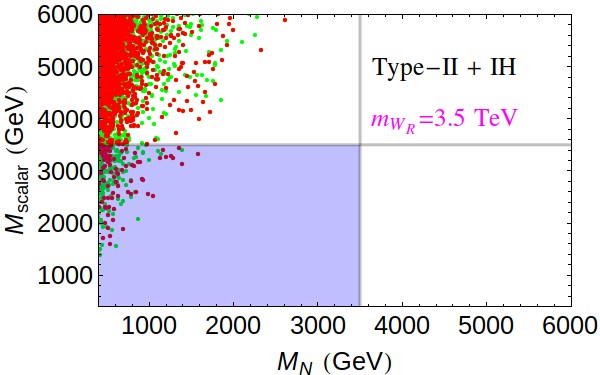}
\caption{The allowed region in the $M_N$ vs $M_{\Delta}$ plane that is 
experimentally allowed by LFV processes, as well as $0 \nu \beta \beta$, for $m_{W_R}=3.5$ TeV. The green points are after satisfying the LFV constraints, while the red points also satisfy the current upper bound on effective mass $m_{ee} < 0.18 ~\rm{eV}$. The blue shaded regions correspond to the natural case with $M_N,~M_{\rm scalar}\leq m_{W_R}$. }
\label{fig:mnmdelta3p5}
\end{figure}

Finally in Figure~\ref{fig:mnmdelta3p5}, we show the allowed region in the $M_N$ versus $M_{\rm scalar}$ plane that is experimentally allowed by both LFV and $0 \nu \beta \beta$ constraints. Here we have set $g_L=g_R$ and $m_{W_R}=3.5$ TeV to satisfy the direct search constraints from the LHC~\cite{Khachatryan:2014dka, Aad:2015xaa}. The blue shaded regions correspond to the case with $M_N,~M_{\rm scalar}\leq m_{W_R}$, which are favoured by vacuum stability and perturbative arguments~\cite{Mohapatra:1986pj, Maiezza:2016bzp}. The green points are after satisfying the LFV constraints, while the red points also satisfy the current upper bound on effective mass $m_{ee} < 0.18 ~\rm{eV}$. It is evident that $M_N$ and $M_{\rm scalar}$ values below 500 GeV or so are disfavoured by the low-energy constraints, whereas heavier triplet masses are still allowed, as long as the corresponding Yukawa couplings are below the perturbative limit.


\section{Diboson excess \label{diboson}}
\begin{figure}[t!]
\subfloat[]{\includegraphics[width=7.1cm, angle =0]{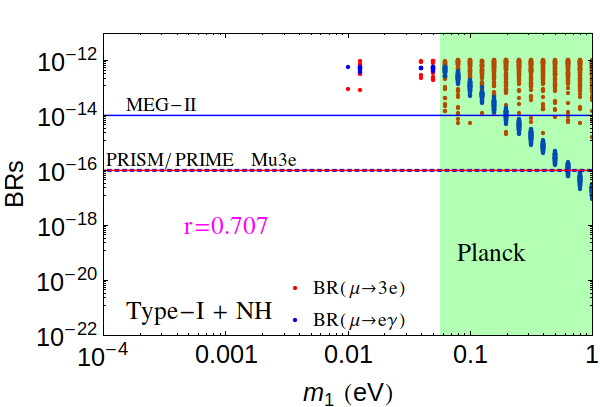}\label{111a}} 
\subfloat[]{\includegraphics[width=7.1cm, angle =0]{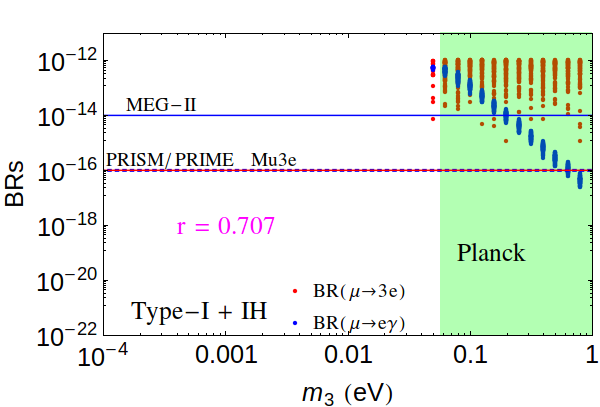}\label{111b}} \\
\subfloat[]{\includegraphics[width=7.1cm, angle =0]{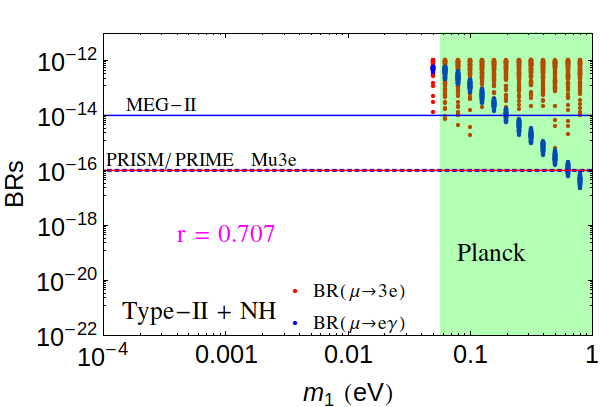}\label{111c}} 
\subfloat[]{\includegraphics[width=7.1cm, angle =0]{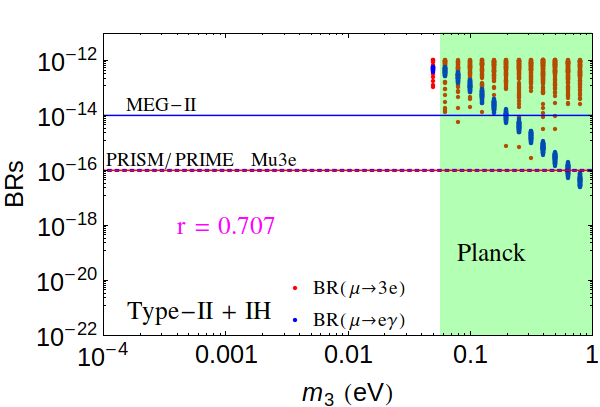}\label{111d}} 
\caption{The branching ratio of $\mu \to e \gamma$ and $\mu \to 3e$ vs light neutrino mass, for the right-handed gauge boson mass $m_{W_R} =$ 2 TeV and  $r = 0.707$.} 
\label{fig:2teva}
\end{figure}
\begin{figure}[t!]
\subfloat[]{\includegraphics[width=7.1cm, angle =0]{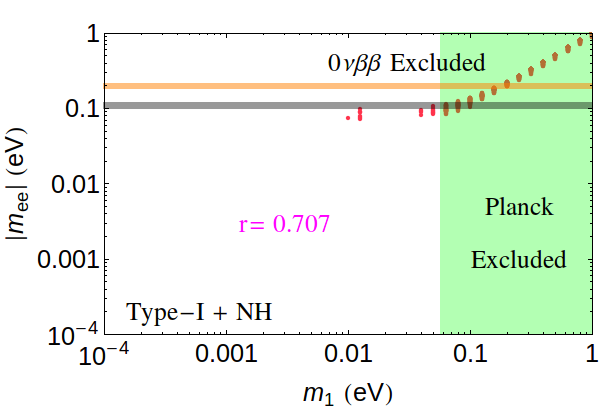}\label{15a}}
\subfloat[]{\includegraphics[width=7.1cm, angle =0]{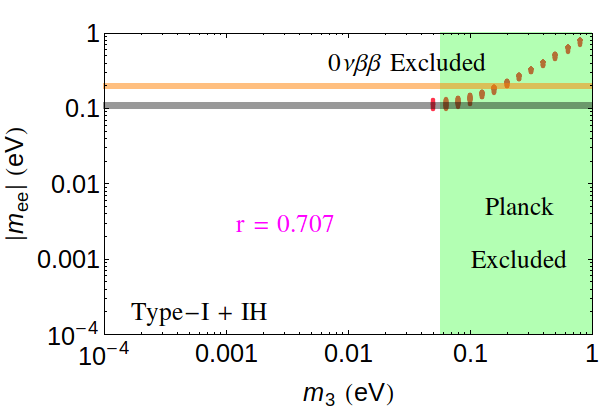}\label{15b}} \\
\subfloat[]{\includegraphics[width=7.1cm, angle =0]{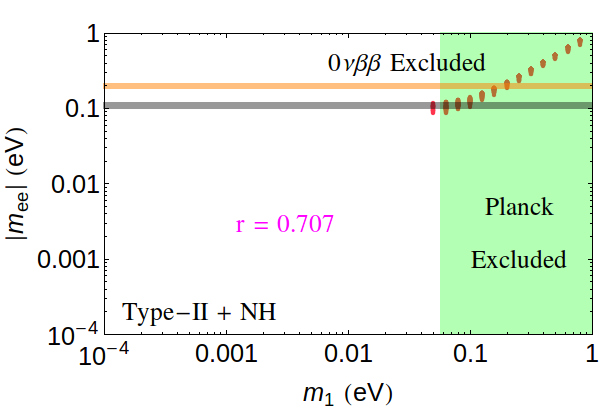}\label{15c}}
\subfloat[]{\includegraphics[width=7.1cm, angle =0]{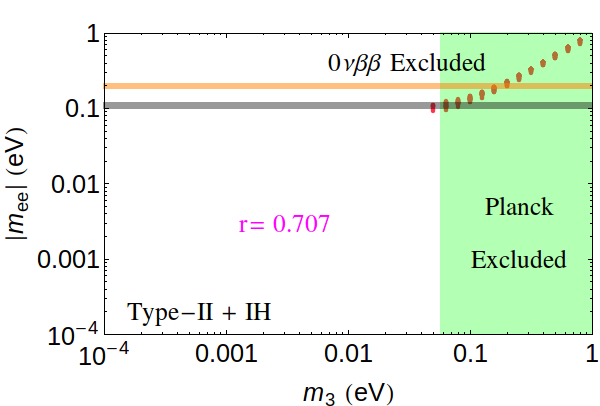}\label{15d}}  
\caption{The effective mass $m_{ee}$  vs light neutrino mass, for the right-handed gauge boson mass $m_{W_R} =$ 2 TeV and  $r = 0.707$. The different panels correspond to:   (a) type-I dominant  NH (b) type-I dominant  IH  (c) type-II dominant  NH (d) type-II dominant  IH. The orange band corresponds to the range of $|m_{ee}| = 0.18 - 0.22$ eV, the region above which is excluded at 90\% C.L. by 
the combined limit from GERDA.
 The black band corresponds to  the future limit ($|m_{ee}| = 0.098 - 0.12$ eV) from GERDA-II. }
\label{fig:2tevb}
\end{figure}

A number of recent resonance searches with the $\sqrt s=8$ TeV LHC data have observed excess events around an invariant mass of 2 TeV, the most notable one being a $3.4\sigma$  local excess in the ATLAS search~\cite{Aad:2015owa, Aad:2015ipg} for a heavy resonance decaying into a pair of SM gauge bosons, followed by the hadronic decay of the diboson system.\footnote{Although no such excess above $2\sigma$ has been found in the early $\sqrt s=13 $ TeV data, the sensitivity is too small to rule out the
Run I excess at the 95\% CL~\cite{ATLAS:2015jj, CMS:2015nmz} and we have to wait for more data from the Run-II phase to confirm/discard this excess.} It is known that this diboson excess can be {\it naturally} explained by a TeV-scale LRSM for the RH gauge boson mass $m_{W_R}\sim 2 $ TeV and the corresponding gauge coupling $g_R\sim 0.4-0.5$~\cite{Hisano:2015gna, Cheung:2015nha,Dobrescu:2015qna,Gao:2015irw,Brehmer:2015cia, Cao:2015lia, Dev:2015pga,Deppisch:2015cua}. 


In this section, we study the implications of the diboson excess on the predictions of LFV and 
$0 \nu \beta \beta$.  A similar study was performed in Ref.~\cite{Awasthi:2015ota}, but here we also include the triplet contribution. For the gauge couplings $g_R \neq g_L$, the branching ratio of the LFV process $\mu \to e \gamma$ is given by Eq.~\eqref{eq:mutoegamma}, where the factors $G^{\gamma}_L$ and $G^{\gamma}_R$ are scaled by a factor of with respect to those given in Eqs.~\eqref{grgamma} and \eqref{glgamma}, {\em i.e.},  
\begin{eqnarray}
 G_R^{\gamma} \ & \simeq & \  \left(\frac{g_R}{g_L}\right)^2\sum_{i}  {V}_{\mu i} {V}^*_{e i} \bigg(\frac{m^2_{W_L}}{m^2_{W_R}}  G_1^{\gamma}(b_i) + \frac{2b_i}{3} \frac{m_{W_L}^2}{m_{\Delta_R^{++}}^2}\bigg),
\label{grgamma2}  \\
 G_L^{\gamma} \ &  \simeq & \ \left(\frac{g_R}{g_L}\right)^2\sum_{i} {V}_{\mu i} {V}^*_{e i} b_i \bigg(\frac{2}{3} \frac{m^2_{W_L}}{m^2_{\Delta_L^{++}}}  G_1^{\gamma}(b_i) + \frac{1}{12} \frac{m_{W_L}^2}{m_{\Delta_L^{+}}^2}\bigg).
\label{glgamma2}
\end{eqnarray}

Similarly, the effective mass for $0\nu \beta \beta$ [{\em c.f.}, Eq.~\eqref{mee}] will be of the following form: 
\begin{eqnarray}
m_{ee} \ = \ \sum_i U_{ei}^2 m_i+ \left(\frac{g_R}{g_L}\right)^4 \langle p^2 \rangle  \left(\frac{m_{W_L}}{m_{W_R}}\right)^4 \bigg ( \sum_i  \frac{V_{ei}^2}{M_i}+ \sum_i\frac{V_{ei}^2 M_i} {m^2_{\Delta_R^{++}}} \bigg).
\end{eqnarray}
In Figures~\ref{fig:2teva} and \ref{fig:2tevb}, we show the branching ratios of $\mu \to e\gamma$, $\mu \to 3e$ processes and the effective mass $m_{ee}$ for the RH gauge boson mass $m_{W_R}=2$ TeV. Comparing Figure~\ref{fig:2teva}  with Figure~\ref{fig:lfvmodr} (for $m_{W_R}=3.5$ TeV), it is evident that 
 even a moderate value of $r=0.707$  is now severely constrained. This is also reflected in Figure~\ref{fig:2tevb} from $0\nu\beta\beta$ limits.


\begin{figure}[t!]
\includegraphics[width=7.1cm, angle =0]{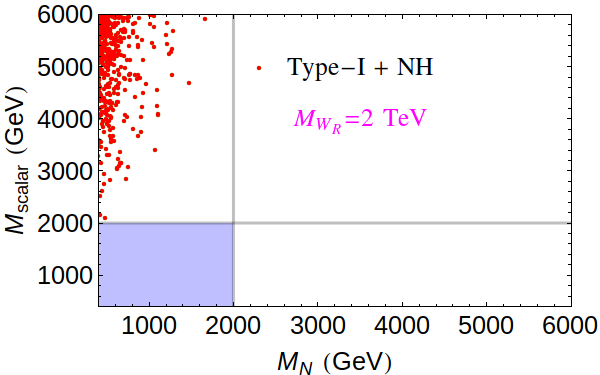}
\includegraphics[width=7.1cm, angle =0]{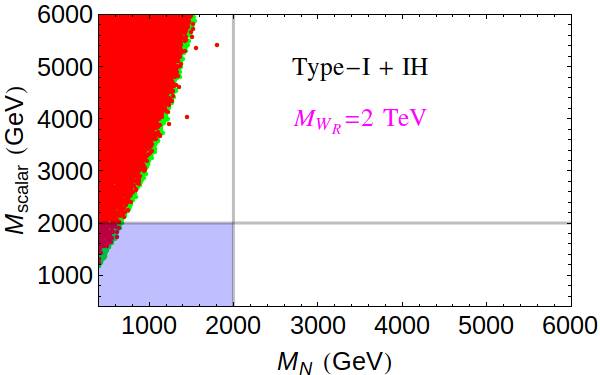}\\
\includegraphics[width=7.1cm, angle =0]{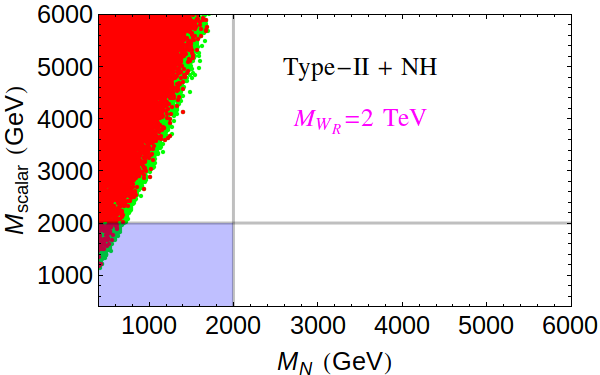}
\includegraphics[width=7.1cm, angle =0]{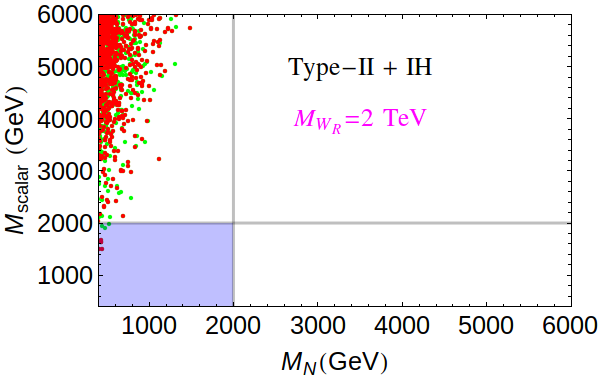}
\caption{The allowed region in the $M_N$ vs $M_{\rm scalar}$ plane that is 
experimentally allowed by LFV processes, as well as $0 \nu \beta \beta$, for $m_{W_R}=2$ TeV. The green points are after satisfying the LFV constraints, while the red points also satisfy the current upper bound on effective mass $m_{ee} < 0.18 ~\rm{eV}$. The blue shaded regions correspond to the natural case with $M_N,~M_{\rm scalar}\leq m_{W_R}$.}
\label{fig:mnmdelta}
\end{figure}

 Finally in Figure~\ref{fig:mnmdelta}, we show the allowed region in the $M_N$ versus $M_{\rm scalar}$ plane that is 
experimentally allowed by LFV processes, as well as by $0 \nu \beta \beta$, while explaining the diboson excess. As in Figure~\ref{fig:mnmdelta3p5}, the blue shaded regions correspond to the natural case with $M_N,~M_{\rm scalar}\leq m_{W_R}$~\cite{Mohapatra:1986pj, Maiezza:2016bzp}. The green points are after satisfying the LFV constraints, while the red points also satisfy the current upper bound on effective mass $m_{ee} < 0.18 ~\rm{eV}$. 

\section{Summary \label{summary}} 
We have studied the correlated constraints from low-energy LFV and $0\nu\beta\beta$ processes for a TeV scale LRSM including the contribution of the Higgs triplets.
Triplet masses comparable to or lighter than the RH neutrino masses were previously thought to be completely ruled out by the LFV constraints. 
We show that even with relatively lower values of triplet masses, it is still possible 
to get allowed parameter regions consistent with the 
LFV limits due to the existence of cancellations between different contributions predominantly in the quasi-degenerate region which can be attributed to the so far unknown $C\!P$ phases.  
We illustrate this effect in a simplified scenario of the LRSM in type-I and type-II seesaw dominance limits for both normal and inverted mass hierarchies, by fixing the RH gauge boson mass $m_{W_R}=3.5$ TeV and the heaviest RH neutrino mass $M_N=500$ GeV, and varying the triplet mass $M_\Delta$ in terms of the ratio $r=M_N/M_\Delta$.  We find that for small values of $r\lesssim 0.1$, the triplet contributions to the LFV observables are negligible (see Figure~\ref{fig:lfvlowr}), in agreement with the previous studies. However, for moderate values of $0.1\lesssim r \lesssim 1$, the triplet contribution  rules out only a part of the LRSM parameter space. In particular,  a hierarchical light neutrino spectrum with $m_1\lesssim 0.01$ eV is
disfavoured  from LFV constraints for type-I NH scenario, while the type-I IH and type-II cases remain largely unconstrained, but can be accessible at future LFV experiments (see Figure~\ref{fig:lfvmodr}). Constraints are also obtained on the Majorana phase $\alpha_2$, restricting it close to either  
0  or $\pi$ for most of the cases analysed here (see Figure~\ref{fig:alpha2}).  
 For larger values of $r\gtrsim 1$, LFV constraints become more stringent, ruling out the hierarchical light neutrino spectrum and only allowing the quasi-degenerate region (see Figure~\ref{fig:lfvhighr}). However, this quasi-degenerate region is already disfavoured by the cosmological limit on the sum of light neutrino masses from Planck data, as well as from current experimental constraints on the half-life of $0\nu\beta\beta$ process. Thus, we conclude that light triplets are completely disfavoured only for $r\gtrsim 1$ (see Figure~\ref{fig:rmsmall}). 

We also give the predictions for the effective neutrino mass for $0\nu\beta\beta$ for all the cases, taking into account the LFV constraints.  Again, we find that for a higher value of $r\gtrsim 1$,  the LRSM parameter space is severely restricted due to the LFV constraints, while the $0\nu\beta\beta$ predictions for moderate values of $0.1\lesssim r \lesssim 1$ are within reach of future experiments (see Figures~\ref{fig:meet1nh}-\ref{fig:meet2ih}).  We emphasise that the LFV 
constraints on the Majorana phases 
play a non-trivial role in ruling out parts of the parameter space otherwise allowed by the $0\nu\beta\beta$ constraints.

Finally, we also study the triplet contribution to LFV and $0\nu\beta\beta$ for the LRSM scenario with $m_{W_R} = 2 ~\rm{TeV}$ and $g_R = 0.5$, being motivated by the recent indication of a diboson excess by the ATLAS experiment. We find that this case is more severely constrained than the $m_{W_R}=3.5$ TeV case discussed above. However, one can find smaller values of $r$ (see Figure~\ref{fig:mnmdelta}) which are still consistent with the LFV and $0\nu\beta\beta$ constraints, while simultaneously explaining the ATLAS diboson anomaly. With more data pouring in from the Run-II phase of the LHC, the light triplet scenario could be probed at the energy frontier in near future, in conjunction with the complementary probes in future low-energy experiments at the intensity frontier.

\section*{Acknowledgments} 
B.D. would like to thank Werner Rodejohann for helpful discussions. This work of B.D. was supported in part by a TUM University Foundation Fellowship and the DFG cluster of excellence ``Origin and Structure of the Universe'', as well as by the DFG with grant RO 2516/5-1 during the final stages. B.D. also acknowledges the local hospitality and partial support from DESY, Hamburg where part of this work was done. M.M. would like to acknowledge the generous support provided by IISER Mohali  and the DST-INSPIRE Faculty Scheme.

\addcontentsline{toc}{section}{References}
\bibliographystyle{JHEP}
\bibliography{BDGMv1}
\end{document}